\documentclass[11pt]{article}
\usepackage{amsmath}
\usepackage{amssymb}
\makeatletter
\usepackage{jheppub}
\usepackage{bbm}
\renewcommand{\[}{\begin{equation}}
\renewcommand{\]}{\end{equation}}
\numberwithin{equation}{section}

\def\CA{{\cal A}}

\def\CD{{\cal D}}
\def\CE{{\cal E}}
\def\CF{{\cal F}}

\def\CI{{\cal I}}
\def\CJ{{\cal J}}
\def\CK{{\cal K}}
\def\CL{{\cal L}}
\def\CR{{\cal R}}

\def\CN{{\cal N}}
\def\CO{{\cal O}}

\def\CS{{\cal S}}
\def\CV{{\cal V}}
\def\CW{{\cal W}}
\def\BC{\mathbb{C}}

\def\BP{\mathbb{P}}
\def\BR{\mathbb{R}}
\def\BZ{\mathbb{Z}}
\def\tr{\text{Tr}}
\def\FR{\mathfrak{R}}

\makeatother

\begin{document}

\preprint{PUPT-2468}
\title{Generalized Indices for $\mathcal{N}=1$ Theories in Four-Dimensions}
\author[a]{Tatsuma Nishioka}
\author[b]{and Itamar Yaakov}

\affiliation[a]{School of Natural Sciences, Institute For Advanced Study,\\ Princeton NJ 08540, USA}
\affiliation[b]{Department of Physics,  Princeton University, Jadwin Hall,\\ Princeton NJ 08544, USA}

\emailAdd{nishioka@ias.edu}
\emailAdd{iyaakov@princeton.edu}

\abstract{We use localization techniques to calculate the Euclidean partition functions for $\mathcal{N}=1$ theories on four-dimensional manifolds $M$ of the form $S^1 \times M_3$, where $M_3$ is a circle bundle over a Riemann surface. 
These are generalizations of the $\CN=1$ indices in four-dimensions including the lens space index.
We show that these generalized indices are holomorphic functions of the complex structure moduli on $M$. We exhibit the deformation by background flat connection.}

\maketitle

\section{Introduction}

Local quantum field theories possess an energy-momentum tensor, a
fact which allows us to consider them on a spacetime with geometry
other than that of Minkowski space. Investigation of the theory on
a compact Euclidean manifold, where even the value of the partition
function can be a meaningful observable, can yield valuable information
about the same theory on flat space. Conversely, we can use our knowledge
of the behavior of the theory on flat space to characterize the manifold.
In either approach, it is usually advantageous to preserve some of
the symmetries of the flat space theory. Specifically, preserving
supersymmetry allows us to take advantage of the attendant simplifications
in the computation of BPS observables, including the partition function.

Four-dimensional Euclidean manifolds preserving rigid supersymmetry
for $\mathcal{N}=1$ theories were considered in \cite{Festuccia:2011ws,Dumitrescu:2012ha,Dumitrescu:2012at,Klare:2012gn}.
The analysis in \cite{Dumitrescu:2012ha} applies to theories which
possess a conserved $U(1)_{R}$ current, in addition to the energy-momentum
tensor. In this work, we use the results of \cite{Dumitrescu:2012ha}
to calculate BPS observables of such theories on a manifold, $M$,
which is the total space of an elliptic fiber bundle over a compact
oriented Riemann surface $\Sigma$. $M$ always has the topology $S^{1}\times M_{3}$,
where $M_{3}$ is a principal $U(1)$-bundle over $\Sigma$. As such,
the supersymmetric partition function on $M$ can be thought of as
a type of super-trace over the Hilbert space of the theory quantized
on the spatial manifold $M_{3}$. Such an object is known as an index.
Familiar examples include the Witten index \cite{Witten:1982df},
where $M$ is the four-torus, and the superconformal index \cite{Kinney:2005ej,Romelsberger:2005eg},
where $M$ is topologically $S^{1}\times S^{3}$. These count, with
appropriate signs and fugacities, the supersymmetric vacua of a theory
and (a subset of) the local BPS operators of a CFT, respectively. The
connection between the partition function on a general $M$ and the
flat space theory is less direct.

Our computational approach is based on localization: a technique which
allows us to reduce a supersymmetry preserving Euclidean path integral
to a smaller integral over the set of fixed points of a supercharge.
For a 4$d$ gauge theory with gauge group $G$, the moduli space of
such fixed points, for our chosen supercharge, will be $\mathcal{M}_{G}^{0}$:
the space of flat $G$-connections on $M$. More generally, the space
of fixed points of a supercharge acting on supersymmetry multiplets
is a \emph{superspace}. The fermionic coordinates are associated with
supersymmetric fermionic modes, those with vanishing action in the
localizing term, in the bosonic background. We will argue that in
the present context these occur only on K{\"a}hler manifolds. $M$
is K{\"a}hler if and only if $M\simeq T^{2}\times\Sigma$. To avoid
dealing with fermionic fixed points, we will therefore further restrict
ourselves to nontrivial circle bundles as $M_{3}$. This rules out the
Witten index. Localization has been applied to the computation of
the superconformal index in \cite{Assel:2014paa,Peelaers:2014ima}
and to the manifold $T^{2}\times S^{2}$ in \cite{Closset:2013sxa}.

The data which parametrizes the Euclidean path integral comes from
the action for the $\mathcal{N}=1$ theory, including background deformations,
the metric on $M$, and other information related to the background
supergravity fields. The general form of the computation shows that
the supersymmetric partition function on $M$ depends only on a finite
subset of these parameters, in agreement with the general results
of \cite{Closset:2013vra,Closset:2014uda}. We will leverage those
results to simplify some of the data from the outset. Specifically,
we avoid choosing a metric altogether and specify the geometry of
$M$ by choosing a complex structure and a holomorphic isometry. We
will then argue for the existence of a compatible metric. Our approach
is similar to the one used to perform localization on Seifert manifolds
in \cite{Kallen:2011ny,Ohta:2012ev}. The result for the partition
function on $M$ should reduce to that of the manifolds considered
in \cite{Ohta:2012ev}, essentially our $M_{3}$ although possibly
with a somewhat restricted metric, in an appropriate limit.

In Section \ref{ss:Setup} we examine the topology and complex structure
of $M$ and review, following \cite{Dumitrescu:2012ha}, how supersymmetry
on $M$ is realized. In Section \ref{ss:SUSY_multiplets}
we discuss the multiplet structure and supersymmetric actions on $M$.
In Section \ref{ss:Localization} we construct the
localizing term and discuss its fixed points. We also set up the computation
of the fluctuation determinants which are then computed using the
equivariant index theorem in Section \ref{ss:IndexTheorem}.
Our final result for the partition function on $M$ of a gauge theory
with gauge group $G$, given in Section \ref{ss:Generalized_Index},
is of the schematic form
\[
Z_{G,r,M_{g,d}}\left(\tau_{\text{cs}},\xi_{\text{FI}},a_{f}\right)=\int_{\mathcal{M}_{G}^{0}\left(g,d\right)}e^{-S_{\text{classical}}\left(\tau_{\text{cs}},\xi_{\text{FI}}\right)}Z_{\text{gauge}}^{g,d}\left(\tau_{\text{cs}}\right)Z_{\text{matter}}^{g,d,r}\left(\tau_{\text{cs}},a_{f}\right) \ ,
\]
where $\tau_{\text{cs}},\xi_{\text{FI}},r$ and $a_{f}$ signify
the dependence on the complex structure, Fayet-Iliopoulos terms, $R$-charges and background flat connections for the flavor symmetry group, respectively. 
The integers $g\ge0$ and $d>0$ are the genus of $\Sigma$
and the first Chern class of $M_{3}\rightarrow\Sigma$. The functions
$Z_{\text{gauge}}$ and $Z_{\text{matter}}$ are the fluctuation determinants
associated with gauge and matter multiplets.

\section{Setup}\label{ss:Setup}

New minimal supergravity can be used to construct supersymmetric actions
for theories with 4$d$ $\mathcal{N}=1$ supersymmetry which have a conserved $U\left(1\right)_R$ symmetry on any Hermitian
four-manifold $M$. The results of \cite{Dumitrescu:2012ha,Klare:2012gn} imply that
on such a manifold one may preserve two supercharges of opposite $R$-charge
if one assumes that the metric on $M$ supports a holomorphic Killing
vector $K$ with holomorphic coefficients. Such a vector represents
a torus isometry acting on $M$, though it may incorporate additional
circle actions under special circumstances. Manifolds of this type
are therefore elliptic fibrations over a Riemann surface. However,
as shown in \cite{Closset:2013vra}, a complex manifold with this
topology does not necessarily support such a vector. As well as constraining
the topology further, we will assume throughout that our manifolds
do.

\subsection{\label{sub:Topology-of-M}Topology of $M$}

We will restrict ourselves to studying the case when $M$ is the
total space of a principal elliptic fiber bundle over a compact oriented
Riemann surface $\Sigma$
\begin{align}
T^2\rightarrow M\xrightarrow{\pi}\Sigma\ .
\end{align}
This is equivalent to requiring that the torus action induced by $K$
is free.\footnote{A free action is one where all isotropy groups are trivial. A less
stringent condition is for a circle (or torus) action to be fixed
point free while having finite isotropy groups. Total spaces with
fixed point free actions are more complicated and will not be considered
here. 
} 
The structure of such a total space has the following classification
(see Corollary 1.5 of \cite{MR1630918})
\begin{enumerate}
\item $M$ is diffeomorphic to $S^{1}\times M_{3}$ where $M_{3}$ is a
principal $U(1)$ bundle 
\begin{align}
S^{1}\rightarrow M_{3}\rightarrow\Sigma\ .
\end{align}
\item The topology of $M$ is completely determined by the genus, $g$,
of the base space $\Sigma$, and the value, $d$, of the first Chern
class of the $U\left(1\right)$ bundle whose total space is $M_{3}$. 
\item $M$ can be constructed as a quotient 
\begin{align}
M=\Theta^{\star}/\left\langle \tau\right\rangle \ ,
\end{align}
by a multiplicative cyclic group generated by a number 
\begin{align}
\tau\in\mathbb{C}^{\star}\ ,\qquad|\tau|>1 \ ,
\end{align}
where $\Theta^{\star}$ is the compliment of the zero section in the
total space of a degree $d$ line bundle on $\Sigma$. 
\item $M$ is K{\" a}hler if and only if $d$ vanishes, in which case it is diffeomorphic
to $T^{2}\times\Sigma$.
\item The integer cohomology of $M$ with $d>0$ is given by
\begin{align}
\begin{aligned}
H^{0}\left(M,\mathbb{Z}\right)&\simeq\mathbb{Z}\ ,\qquad H^{1}\left(M,\mathbb{Z}\right)\simeq\mathbb{Z}^{2g+1}\ ,\qquad H^{2}\left(M,\mathbb{Z}\right)\simeq\mathbb{Z}_{d}\oplus\mathbb{Z}^{4g} \ ,\\
&H^{3}\left(M,\mathbb{Z}\right)\simeq\mathbb{Z}_{d}\oplus\mathbb{Z}^{2g+1}\ ,\qquad H^{4}\left(M,\mathbb{Z}\right)\simeq\mathbb{Z}\ ,
\end{aligned}
\end{align}
such that
\begin{align}
\text{Tor}\left(H^{2}\left(M,\mathbb{Z}\right)\right)=\pi^{*}\left(H^{2}\left(\Sigma,\mathbb{Z}\right)\right)\simeq\mathbb{Z}_{d} \ .
\end{align}

\end{enumerate}
We will restrict ourselves mostly to the case $d>0$.

\subsection{Supersymmetry on $M$}

This section is a review of the relevant facts about Killing spinors and vectors on a Hermitian manifold $M$ from \cite{Dumitrescu:2012ha}. Our conventions, which differ somewhat from \cite{Dumitrescu:2012ha}, are summarized in Appendix \ref{ss:Convention}. We begin by discussing the general class of manifolds which admit two Killing spinors of opposite $R$-charge, then specialize to the fiber bundles described in the previous section. The question of finding an appropriate metric on these spaces is deferred until the end of Section \ref{ss:CSandRsym}.

\subsubsection{Killing spinors and spinor bilinears}

The Killing spinor equations on $M$ are read off from the variation
of the gravitinos of the new minimal supergravity \cite{Sohnius:1981tp}
\begin{align}
\begin{aligned}\delta\psi_{\mu} & =\left(\nabla_{\mu}-i\left(A_{\mu}-V_{\mu}\right)-iV^{\nu}\sigma_{\mu\nu}\right)\epsilon=0\ ,\\
\delta\tilde{\psi}_{\mu} & =\left(\nabla_{\mu}+i\left(A_{\mu}-V_{\mu}\right)+iV^{\nu}\bar{\sigma}_{\mu\nu}\right)\tilde{\epsilon}=0\ ,
\end{aligned}
\label{KSNM}
\end{align}
where $\epsilon$ and $\tilde{\epsilon}$ are Killing spinors of $R$-charge
$1$ and $-1$. Namely, $\epsilon$ and $\tilde{\epsilon}$
are sections of $L\otimes S_{+}$ and $L^{-1}\otimes S_{-}$, respectively,
where $L$ is an $R$-symmetry line bundle, $S_{+}$ a left-handed spinor
bundle and $S_{-}$ a right-handed spinor bundle.

The background fields $A_{\mu}$ and $V_{\mu}$ are complex in general.
The real part of $A_{\mu}$ is the connection on the $R$-symmetry line
bundle $L$. $V_{\mu}$ is a conserved current 
\begin{align}
\nabla_{\mu}V^{\mu}=0\ .
\end{align}
The complex conjugated spinor $\epsilon^{\dagger}$ ($\tilde{\epsilon}^{\dagger}$)
satisfies the same Killing spinor equation as $\epsilon$ ($\tilde{\epsilon}$)
upon replacing $A_{\mu}\to-\bar{A}_{\mu}$ and $V_{\mu}\to-\bar{V}_{\mu}$
in \eqref{KSNM}.

The Killing spinor equations on $M$ preserve two supercharges of
opposite $R$-charge and handedness $\epsilon,\tilde{\epsilon}$. They
have the property that everywhere on $M$ their norm do not vanish
\begin{align}
|\epsilon|^{2}\ne0\ ,\qquad |\tilde{\epsilon}|^{2}\ne0\ .
\end{align}
We will regard $\epsilon$ and $\tilde{\epsilon}$ as commuting spinors
below. 

We can use the spinors $\epsilon,\tilde{\epsilon}$ to define real,
(anti-)self-dual two-forms 
\begin{align}
J_{\mu\nu}=-\frac{2i}{|\epsilon|^{2}}\epsilon^{\dagger}\sigma_{\mu\nu}\epsilon\ ,\qquad\tilde{J}_{\mu\nu}=-\frac{2i}{|\tilde{\epsilon}|^{2}}\tilde{\epsilon}^{\dagger}\bar{\sigma}_{\mu\nu}\tilde{\epsilon}\ ,
\end{align}
which are integrable almost complex structures
\footnote{It follows from the Fiertz identity of commuting spinors
\begin{align}
\begin{aligned}(\epsilon_{1}\epsilon_{2})(\tilde{\epsilon}_{3}\tilde{\epsilon}_{4}) & =\frac{1}{2}(\epsilon_{1}\sigma^{\mu}\tilde{\epsilon}_{4})(\epsilon_{2}\sigma_{\mu}\tilde{\epsilon}_{3})\ .
\end{aligned}
\end{align}
The rest of the relations also follow from this identity.
} \cite{Dumitrescu:2012ha,Dumitrescu:2012at} 
\begin{align}
J_{~\rho}^{\mu}J_{~\nu}^{\rho}=\tilde{J}_{~\rho}^{\mu}\tilde{J}_{~\nu}^{\rho}=-\delta_{~\nu}^{\mu}\ .
\end{align}
The two-forms defined by 
\begin{align}
P_{\mu\nu}=\epsilon\sigma_{\mu\nu}\epsilon\ ,\qquad\tilde{P}_{\mu\nu}=\tilde{\epsilon}\bar{\sigma}_{\mu\nu}\tilde{\epsilon}\ ,
\end{align}
satisfy the relations 
\begin{align}
J_{\mu}^{~\rho}P_{\rho\nu}=iP_{\mu\nu}\ ,\qquad\tilde{J}_{\mu}^{~\rho}\tilde{P}_{\rho\nu}=i\tilde{P}_{\mu\nu}\ .
\end{align}
$P_{\mu\nu}$ ($\tilde{P}_{\mu\nu}$) is a section of $L^{2}\otimes\Lambda_{+}^{2}$
($L^{-2}\otimes\Lambda_{-}^{2}$). $\Lambda_{+}^{2}$ ($\Lambda_{-}^{2}$)
is the bundle of (anti-)self-dual two-forms.

One can also construct a vector field by combining $\epsilon$ and
$\tilde{\epsilon}$ 
\begin{align}
K^{\mu}=\epsilon\sigma^{\mu}\tilde{\epsilon}\ .
\end{align}
It follows from \eqref{KSNM} that $K^{\mu}$ is a holomorphic Killing
vector 
\begin{align}\label{KillingVector}
\begin{aligned}
\nabla_{\mu}K_{\nu}+\nabla_{\nu}K_{\mu}&=0\ , \qquad
J^{\mu}_{~\nu}K^{\nu}={\tilde{J}^{\mu}}_{~\nu}K^{\nu}=iK^{\mu} \ ,
\end{aligned}
\end{align}
and hence
\begin{align}
\begin{aligned}
K^{\dagger\mu}K_{\mu}&\ne0\ , &\qquad K^{\mu}K_{\mu}&=0 \ , \\
|\text{Re}\left(K\right)|^{2}=|\text{Im}\left(K\right)|^{2}&\ne0\ ,&\qquad\text{Re}\left(K\right)^{\mu}\text{Im}\left(K\right)_{\mu}&=0 \ .
\end{aligned}
\end{align}

We will restrict attention to the generic case where $K$ commutes
with its conjugate
\[
\left[K,K^{\dagger}\right]=0\ ,
\]
This is enough to show that $M$ is a torus fibration, with a torus
isometry action induced by the real and imaginary parts of $K$, over
a Riemann surface $\Sigma$ \cite{Dumitrescu:2012ha}. We will take
the torus action to be free and hence $M$ is the total space of a
principal torus bundle. The orbits of $K$ need not close, but may
be part of a larger $U(1)^{3}$ group of isometries \cite{Closset:2013vra}.
Since the metric on $M$ is, by definition, constant along the fibers
(though the size of the fibers can vary with position on the base),
an extra $U(1)$ implies that there exists a Killing vector for the
quotient metric on $\Sigma$. Riemann surfaces supporting such a metric
exist only for $g\le1$. The extra Killing vector is unique (up to
rescaling) for $g=0$ and is one of the two translations of the torus, or a linear combination thereof, for $g=1$. Note that fixed points
for the action of such vectors exist only for $g=0$.

We also introduce independent vectors
\footnote{Note that $K^{\dagger}\ne\bar{K}$ in general, however the two can be made equal using a conformal transformation of the metric. We hope this will not cause too
much confusion.
} 
\begin{align}
\bar{K}^{\mu}=-\zeta\sigma^{\mu}\tilde{\zeta}\ ,\qquad Y^{\mu}=\epsilon\sigma^{\mu}\tilde{\zeta}\ ,\qquad\bar{Y}^{\mu}=\zeta\sigma^{\mu}\tilde{\epsilon}\ ,\label{DefY}
\end{align}
where $\zeta$ and $\tilde{\zeta}$ are defined by 
\begin{align}\label{ZetaSpinor}
\zeta\equiv\frac{\epsilon^{\dagger}}{|\epsilon|^{2}}\ ,\qquad\tilde{\zeta}\equiv\frac{\tilde{\epsilon}^{\dagger}}{|\tilde{\epsilon}|^{2}}\ ,\qquad\zeta\epsilon=\tilde{\zeta}\tilde{\epsilon}=1\ .
\end{align}
These vectors satisfy the following: 
\begin{align}\label{KYrelation}
\begin{aligned}K^{\mu}\bar{K}_{\mu} & =Y^{\mu}\bar{Y}_{\mu}=2\ ,\\
K^{\mu}K_{\mu}=\bar{K}^{\mu}\bar{K}_{\mu}=Y^{\mu}Y_{\mu}=\bar{Y}^{\mu}\bar{Y}_{\mu} & =K^{\mu}Y_{\mu}=K^{\mu}\bar{Y}_{\mu}=\bar{K}^{\mu}Y_{\mu}=\bar{K}^{\mu}\bar{Y}_{\mu}=0\ ,
\end{aligned}
\end{align}
The metric, complex structures and two-forms are written in terms of the vectors as
\begin{align}\label{KYdecomposition}
\begin{aligned}
g_{\mu\nu} & =\frac{1}{2}\left(K_{\mu}\bar{K}_{\nu}+K_{\nu}\bar{K}_{\mu}+Y_{\mu}\bar{Y}_{\nu}+Y_{\nu}\bar{Y}_{\mu}\right)\ ,\\
J_{\mu\nu} & =\frac{i}{2}\left(K_{\mu}\bar{K}_{\nu}-K_{\nu}\bar{K}_{\mu}+Y_{\mu}\bar{Y}_{\nu}-Y_{\nu}\bar{Y}_{\mu}\right)\ ,\\
\tilde{J}_{\mu\nu} & =\frac{i}{2}\left(K_{\mu}\bar{K}_{\nu}-K_{\nu}\bar{K}_{\mu}-Y_{\mu}\bar{Y}_{\nu}+Y_{\nu}\bar{Y}_{\mu}\right)\ ,\\
P_{\mu\nu} & =\frac{1}{2}\left(K_{\mu}Y_{\nu}-K_{\nu}Y_{\mu}\right)\ ,\\
\tilde{P}_{\mu\nu} & =\frac{1}{2}\left(K_{\mu}\bar{Y}_{\nu}-K_{\nu}\bar{Y}_{\mu}\right)\ .
\end{aligned}
\end{align}
It follows that $K,Y$ ($\bar{K},\bar{Y}$) are (anti-)holomorphic vectors. 

We can decompose arbitrary spinors as
\begin{align}\label{eq:left_handed_decomposition}
\begin{aligned}
\psi_{\alpha} &=\left(\zeta\psi\right)\epsilon_{\alpha}-\left(\epsilon\psi\right)\zeta_{\alpha}\ , \\
\tilde{\psi}_{\dot{\alpha}} &=\left(\tilde{\zeta}\tilde{\psi}\right)\tilde{\epsilon}_{\dot{\alpha}}+\left(\tilde{\epsilon}\tilde{\psi}\right)\tilde{\zeta}_{\dot{\alpha}} \ ,
\end{aligned}
\end{align}
from which we can recover
\begin{align}
\begin{aligned}
\tilde{\epsilon}\bar{\sigma}^{\mu}\lambda &=\left(\zeta\lambda\right)K^{\mu}+\left(\epsilon\lambda\right)Y^{\mu} \ ,\\
\epsilon\sigma^{\mu}\tilde{\lambda}&=\left(\tilde{\zeta}\tilde{\lambda}\right)K^{\mu}-\left(\tilde{\epsilon}\tilde{\lambda}\right)\bar{Y}^{\mu} \ .
\end{aligned}
\end{align}

\subsubsection{The complex manifold $M$}

Introducing complex coordinates $w,z$ such that $K=\partial_{w}$,
the metric on $M$ can be written as 
\begin{align}
ds^{2}=\Omega(z,\bar{z})^{2}\left((dw+h(z,\bar{z})dz)(d\bar{w}+\bar{h}(z,\bar{z})d\bar{z})+c(z,\bar{z})^{2}dzd\bar{z}\right)\ .\label{T2OverRS}
\end{align}
The Hermitian manifold $M$ admits a Chern connection that is compatible
with the metric and the complex structure 
\[
\nabla_{\mu}^{c}g_{\nu\rho}=0\ ,\qquad\nabla_{\mu}^{c}J_{\nu\rho}=0\ .
\]
The second condition is equivalent to 
\begin{align}
\nabla_{\mu}J_{\nu\rho}-(\Gamma^{c})_{[\mu\nu]}^{\sigma}J_{\sigma\rho}-(\Gamma^{c})_{[\mu\rho]}^{\sigma}J_{\nu\sigma}=0\ ,
\end{align}
where $(\Gamma^{c})_{[\mu\nu]}^{\sigma}$ is the Christoffel symbol
whose lower indices are anti-symmetrized. Rotating the three indices
and taking an appropriate summation, one obtains the Christoffel tensor represented by the complex structure
\begin{align}
(\Gamma^{c})_{[\mu\nu]}^{\sigma}=\frac{1}{2}J^{\rho\lambda}(\nabla_{\mu}J_{\nu\lambda}+\nabla_{\lambda}J_{\mu\nu}-\nabla_{\nu}J_{\lambda\mu})\ .
\end{align}
Since the symmetric part is the usual Christoffel symbol of the Levi-Civita
connection, the spin connection takes the form 
\begin{align}
\begin{aligned}(\omega^{c})_{\mu}^{mn} & =e_{\nu}^{m}\left(\partial_{\mu}e^{\nu n}+(\Gamma^{c})_{\mu\sigma}^{\nu}e^{\sigma n}\right) \ ,\\
 & =\omega_{\mu}^{mn}+\frac{1}{2}e_{\rho}^{n}e^{\nu n}J^{\rho\lambda}(\nabla_{\mu}J_{\nu\lambda}+\nabla_{\lambda}J_{\mu\nu}-\nabla_{\nu}J_{\lambda\mu})\ .
\end{aligned}
\end{align}

Now we rewrite the Killing spinor equation \eqref{KSNM} by using
the Chern connection 
\begin{align}
\begin{aligned}(\nabla_{\mu}^{c}-iA_{\mu}^{c})\epsilon & =0\ ,\end{aligned}
\end{align}
where we defined 
\begin{align}
A_{\mu}^{c}=A_{\mu}+\frac{1}{4}(\delta_{\mu}^{\nu}-iJ_{\mu}^{~\nu})\nabla_{\rho}J_{~\nu}^{\rho}-\frac{3}{2}\kappa K_{\mu}\ .
\end{align}
$\kappa$ is an undetermined scalar function satisfying 
\[
K^{\mu}\partial_{\mu}\kappa=0\ .
\]
To determine the connection $A_{\mu}^{c}$ in terms of the Chern connection,
consider $p=\tilde{P}_{12}\in L^{-2}\otimes\CK_{M}$, where $\CK_{M}=\Lambda^{2,0}$
is the canonical bundle of $(2,0)$-forms. Since $p$ is a bilinear
of two Killing spinors, it satisfies 
\begin{align}
(\nabla_{\mu}^{c}+2iA_{\mu}^{c})p & =0\ .\label{pEquation}
\end{align}
The fact that $p$ is globally well-defined implies the line bundle
$L^{-2}\otimes\mathcal{K}_{M}$ is topologically trivial. Also, the
fact that the Christoffel symbols of mixed indices with and without
bar vanish leads to 
\begin{align}\label{RChern}
\nabla_{i}^{c}p=\partial_{i}p-\frac{p}{2}\partial_{i}\log g\ ,\qquad\nabla_{\bar{i}}^{c}p=\partial_{\bar{i}}p\ ,
\end{align}
where we used $g\equiv\det g_{\mu\nu}=(\det g_{i\bar{j}})^{2}$. 
Using them in \eqref{pEquation}, we obtain 
\[
A^{c}=-\frac{i}{8}\left(\partial-\bar{\partial}\right)\log\, g+\frac{i}{2}\left(\partial+\bar{\partial}\right)\log\, s \ ,
\]
where $s\equiv pg^{-1/4}$ is a nowhere vanishing function. The $R$-symmetry gauge field
is alternatively given by \cite{Closset:2013sxa} 
\[
A_{\mu}=-\frac{1}{4}{J_{\mu}}^{\nu}\partial_{\nu}\log\,\sqrt{g}-\frac{1}{4}\left({\delta^{\nu}}_{\mu}-i{J_{\mu}}^{\nu}\right)\nabla_{\rho}{J^{\rho}}_{\nu}+\frac{i}{2}\partial_{\mu}\log\, s+\frac{3}{2}\kappa K_{\mu}\ .
\]
Note that
\[
A_{w}=\frac{i}{2}\partial_{w}\log\, s \ .
\]

Since the Ricci form is given by 
\begin{align}
\CR=i\partial\bar{\partial}\log\sqrt{g}=-\frac{i}{4}d(\partial-\bar{\partial})\log g\ ,
\end{align}
the field strength of $A_{\mu}^{c}$ is proportional
to the first Chern class $c_{1}(M)=\left[\frac{\CR}{2\pi}\right]$
up to an exact two-form \cite{Cassani:2013dba} 
\begin{align}
\left[F^{(A^{c})}\right]=\pi\left[c_{1}(M)\right]\ .\label{FirstChern}
\end{align}

We may choose the vielbein in the Hermitian coordinates \eqref{T2OverRS} to be 
\begin{align}
e^{1}=\Omega(dw+hdz)\ ,\qquad e^{2}=\Omega\, c\,dz\ ,
\end{align}
leading to Killing spinors of the form 
\begin{align}\label{KSsolutions}
\epsilon_{\alpha}=\frac{1}{\sqrt{s}}\left(\begin{array}{c}
0\\
1
\end{array}\right)\ ,\qquad\tilde{\epsilon}^{\dot{\alpha}}= \frac{\sqrt{s}\,\Omega}{2} \left(\begin{array}{c}
0\\
1
\end{array}\right)\ .
\end{align}
Using the Killing spinors \eqref{KSsolutions} and the sigma matrices in the Hermitian coordinates \eqref{SigmaMatHermite}, the vectors become
\begin{align}\label{KillingVectors}
\begin{aligned}
K &= \partial_{w} \  ,\\
\bar K &=  \frac{4}{\Omega^2} \partial_{\bar w} \ , \\
Y &= \frac{4}{\Omega^2 c s}\left(\partial_{z}-h\partial_{w} \right)\ , \\
\bar Y &=\frac{s}{c}\left(\partial_{\bar{z}}-\bar{h}\partial_{\bar{w}}\right) \ .
\end{aligned}
\end{align}
At this point we set $\Omega =2$ to have $K^\dagger = \bar K$ for simplicity. This choice of $\Omega$ is irrelevent for the computation of the partition function.

\subsubsection{The integrability conditions}

The condition $[\nabla_{\mu},\nabla_{\nu}]\epsilon=\frac{1}{2}R_{\mu\nu\rho\sigma}\sigma^{\rho\sigma}\epsilon$
and the Killing spinor equation yields the integrability condition
\begin{align}
\begin{aligned}\frac{1}{2}R_{\mu\nu\rho\sigma}\sigma^{\rho\sigma}\epsilon & =-V^{\rho}V_{\rho}\sigma_{\mu\nu}\epsilon+i\left(F_{\mu\nu}^{A}-F_{\mu\nu}^{V}\right)\epsilon\\
 & \qquad+i\left(\nabla_{\mu}+iV_{\mu}\right)V^{\rho}\sigma_{\nu\rho}\epsilon-i\left(\nabla_{\nu}+iV_{\nu}\right)V^{\rho}\sigma_{\mu\rho}\epsilon\ ,
\end{aligned}
\end{align}
where $F^{A,V}_{\mu\nu}$ are the field strength of the vector fields $A_\mu,V_\mu$.
Contracting both sides with $\epsilon^{\dagger}\sigma^{\mu\nu}$ from
the left, we obtain 
\begin{align}
R-6V^{\mu}V_{\mu}=-2F_{\mu\nu}^{A}J^{\mu\nu}\ ,\label{IntegCond}
\end{align}
while a contraction with $\epsilon\sigma^{\mu\nu}$ yields 
\begin{align}
F_{\mu\nu}^{A}P^{\mu\nu}=0\ .
\end{align}
Similarly, the integrability condition for $\tilde{\epsilon}$ gives
equalities 
\begin{align}
R-6V^{\mu}V_{\mu}=2F_{\mu\nu}^{A}\tilde{J}^{\mu\nu}\ ,\qquad F_{\mu\nu}^{A}\tilde{P}^{\mu\nu}=0\ .\label{IntegCondBar}
\end{align}

The integrability conditions for two supercharges with opposite $R$-charge
leads to other interesting relations between the space-time curvatures
and the background field strengths. They are given in the following
forms \cite{Cassani:2013dba} 
\begin{align}
\begin{aligned}(C_{\mu\nu\rho\sigma})^{2} & =\frac{8}{3}\text{Re}\,(\CF_{\mu\nu})^{2}\ ,\\
\epsilon^{\mu\nu\rho\sigma}R_{\mu\nu\alpha\beta}R_{\rho\sigma}^{~~\alpha\beta} & =\frac{8}{3}\text{Re}\left[\epsilon^{\mu\nu\rho\sigma}\CF_{\mu\nu}\CF_{\rho\sigma}\right]\ ,\\
\text{Im}\,(\CF_{\mu\nu})^{2} & =\text{Im}\left[\epsilon^{\mu\nu\rho\sigma}\CF_{\mu\nu}\CF_{\rho\sigma}\right]=0\ ,
\end{aligned}
\label{WeylAndPontryagin}
\end{align}
where $\CF$ is the field strength of the background $U(1)$ gauge
field 
\begin{align}
\CF=d\CA\ ,\qquad\CA_{\mu}=A_{\mu}-2V_{\mu}\ .
\end{align}

\subsection{Complex structure and $R$-symmetry background}\label{ss:CSandRsym}

The supergravity background may require including an $R$-symmetry
background on $M$, possibly incorporating a nontrivial line bundle
$L$. The condition given in \cite{Dumitrescu:2012ha} is that $L^{-2}\times\mathcal{K}_{M}$
is trivial, where $\mathcal{K}_{M}$ is the canonical line bundle
on $M$. We will determine the topological class of $L$ when we examine
the complex structure. Note that the $R$-symmetry gauge field, $A$,
is in general complex. However, only the real part of $A$ can be
a connection for a nontrivial line bundle.

For $g\ge1$ and $d\ge1$ all complex structures of $M$ are deformation
equivalent \cite{nakagawa1995complex}. The canonical bundle is a
pullback from the base \cite{hofer1993remarks} 
\[
\mathcal{K}_{M}=\pi^{\star}\mathcal{K}_{\Sigma}\ ,
\]
and hence satisfies \cite{MR1630918} 
\[
c_{1}\left(\mathcal{K}_{M}\right)=\pi^{\star}c_{1}\left(\mathcal{K}_{\Sigma}\right)=2g-2~\text{mod }d \in\mathbb{Z}_{d}\subset H^{2}\left(M,\mathbb{Z}\right)\ .
\]
We also have
\[
c_{1}\left(\mathcal{K}_{M}\right)\,\text{mod}\:2=w_{2}\left(T_{M}\right)=0\in H^{2}\left(M,\mathbb{Z}_{2}\right)\ ,
\]
where $w_{2}\left(T_{M}\right)$ is the second Stiefel-Whitney
class of $M$.

The condition on the $R$-symmetry line bundle yields 
\[
-2c_{1}\left(L\right)+2g-2=0~\text{mod }d\ .
\]
We will ignore 2-torsion and only consider the solution 
\begin{align}
c_{1}\left(L\right)=g-1~\text{mod }d\ .\label{eq:R_symmetry_Chern_class}
\end{align}

There is a subtlety associated with the case of $g=0$. The total
space is then diffeomorphic to 
\[
M\simeq S^{1}\times L\left(d,1\right)\ ,
\]
where $L\left(r,s\right)$ is a (three-dimensional) lens space. For
$d\ge3$ the complex structure moduli space of $M$ has two deformation
equivalence classes $\text{I,II}$ \cite{nakagawa1995complex}. From
the fact that the usual lens space index has a (topologically) trivial
$R$-symmetry bundle, we conclude that the topological classification
of the canonical bundle in this case is 
\[
\mathcal{K}_{M}=\begin{cases}
\text{topologically trivial} & \text{I} \ ,\\
\text{\ensuremath{\pi^{\star}\mathcal{K}_{\Sigma}}} & \text{II} \ ,
\end{cases}
\]
so that 
\[
c_{1}\left(\mathcal{K}_{M}\right)=\begin{cases}
0 & \text{I} \ ,\\
-2\in\mathbb{Z}_{d} & \text{II} \ ,
\end{cases}
\]
and our solution for the $R$-symmetry line bundle is 
\[
c_{1}\left(L\right)=\begin{cases}
0 & \text{I} \ ,\\
-1\in\mathbb{Z}_{d} & \text{II} \ .
\end{cases}
\]
An example of this phenomenon is that the spaces $S^{1}\times L\left(d,1\right)$
and $S^{1}\times L\left(d,-1\right)$ are diffeomorphic but have,
in the language of \cite{Closset:2013vra}, topologically distinct
canonical bundles for $d\ge3$ with first Chern classes given by the
two solutions above. 
In order to have chiral fields valued in well-defined line bundles, one must make the following restriction on the $R$-charges
\[
r\left(-\frac{\chi\left(\Sigma\right)}{2}\text{ mod }d\right)\in\mathbb{Z}\ .
\]
For instance, in the complex structure $\text{II}$, the $R$-charges are quantized in units of $d-1$. Note that there is no restriction for the complex structure of type $\text{I}$, and for manifolds with base space $T^2$. The case $g=0$ and $d=0$, where $M$ is diffeomorphic to $T^2\times S^2$ and which we consider only briefly, is very different. One is forced to include flux for the $R$-symmetry gauge field on $S^2$ and all chiral fields must have integer $R$ charges. 

In order to use the results about supersymmetry from the previous section, we must show that $M$ admits a compatible Hermitian metric which supports $K$. When $g\ge2$ the orbits of $K$ are tori and the fibration is holomorphic
\cite{Closset:2013sxa}. We take this to mean that a metric with an
appropriate Killing vector can be constructed on $M$ by averaging
any Hermitian metric along the fibers. The Killing vector $K$, which
simply points along the fiber directions, should be holomorphic in
the given complex structure. For $g=0$, appropriate metrics were constructed in  \cite{Closset:2013vra}. It should be noted that there is no guarantee that a complex manifold
with $g=0$ and a complex structure of type $\text{II}$ will admit the necessary
holomorphic Killing vector.\footnote{The combined results of \cite{nakagawa1995complex} and \cite{Closset:2013vra}
do not seem to rule out this possibility. That is, one may take $q=1$
and $r=\pm1$ with $\lambda=0$, in the language of \cite{nakagawa1995complex},
and the metric and Killing vector $(4.7)$ and $(4.8)$ respectively from
\cite{Closset:2013vra}. In the language of \cite{Closset:2013vra}
the options correspond to $s=\pm1$ which produce the correct Chern
classes, assuming the formulas there hold beyond their specified region
$1\le s<r$. 
}
We currently have nothing to say about the case $g=1$. 

Note that even when the $R$-symmetry line bundle is determined to
be topologically trivial it may be holomorphically nontrivial. We
will need to evaluate the determinant of the gauge invariant operator
\[
\delta_{K}=\mathcal{L}_{K}-irK^{\mu}A_{\mu}-iq_{f}K^{\mu}a_{\mu}\ ,
\]
and note that 
\[
K^{\mu}A_{\mu}=A_{w}\ ,
\]
is a holomorphic line bundle modulus for $A$. For the cases where
$\mathcal{K}_{M}$ is a pullback from the base, the modulus on the
base is given by $1/2$ that of the base canonical bundle. We know
from the explicit form of the lens space partition function that this
remains true even in the special component of the complex structure
moduli space ($\text{I}$) given above. This is due to the fact that
one can reach this component by orbifolding $S^{3}\times S^{1}$,
to which the above argument applies, without changing the other supergravity
fields \cite{Closset:2013vra}. We do not explicitly include a holonomy for $A$
around the $S^{1}$, but treat all spinors as periodic. The holonomy
in the fiber direction is constrained by the topological class of
the $R$-symmetry bundle. As explained in Section \ref{sub:The-bosonic-moduli-space},
the solution \eqref{eq:R_symmetry_Chern_class} for the Chern class
of $L$ implies a holonomy in the fiber direction of size 
\[\label{R_bundle_holonomy}
\exp\left(2\pi i\frac{c_{1}\left(L\right)}{d}\right)=\exp\left(2\pi i\frac{g-1}{d}\right)\ .
\]
This holonomy is not included when working with the alternative complex
structure I.

\section{$\CN =1$ supersymmetry algebra and multiplets}\label{ss:SUSY_multiplets}

The supersymmetry transformations of the new minimal supergravity \cite{Sohnius:1981tp}
satisfy the following commutation relations in the rigid limit \cite{Festuccia:2011ws,Dumitrescu:2012ha}
\begin{align}\label{SUSY_Alg}
\begin{aligned}\{\delta_{\epsilon},\delta_{\tilde{\epsilon}}\} & =\frac{1}{2}\delta_{K}\ ,\\
\{\delta_{\epsilon},\delta_{\epsilon}\} & =\{\delta_{\tilde{\epsilon}},\delta_{\tilde{\epsilon}}\}=0\ ,\\
 & =[\delta_{K},\delta_{\epsilon}]=0\ ,\\
 & =[\delta_{K},\delta_{\tilde{\epsilon}}]=0\ ,
\end{aligned}
\end{align}
where $\delta_{\epsilon}$ and $\delta_{\tilde{\epsilon}}$ are the
supersymmetry transformations with respect to supercharges $\epsilon$
and $\tilde{\epsilon}$, respectively. $\delta_{K}$ is the $R$-covariant
Lie derivative 
\begin{align}
\delta_{K}=\CL_{K}^{A}=\CL_{K}-ir\, K^{\mu}A_{\mu}\ ,
\end{align}
and $\CL_{K}$ is the Lie derivative along the Killing vector $K^{\mu}$ given
by \eqref{KillingVector}. The equalities in the third line of \eqref{SUSY_Alg}
follow from the fact that Killing spinors are $R$-covariantly constant
along the Killing vector $K$ 
\begin{align}
\CL_{K}^{A}\,\epsilon=\CL_{K}^{A}\,\tilde{\epsilon}=0\ .\label{LieKS}
\end{align}

The most general multiplet, denoted by $\CS$, whose transformation law realizes
the supersymmetry algebra \eqref{SUSY_Alg} consists of $16+16$ bosonic
and fermionic degrees of freedom \cite{Sohnius:1982fw,Kugo:1982cu,Closset:2012ru}
\begin{align}
\CS=(C,\chi,\tilde{\chi},M,\tilde{M},a_{\mu},\lambda,\tilde{\lambda},D)\ .
\end{align}
If the bottom component $C$ has $R$-charge $r$, $a_{\mu}$ and
$D$ have the same $R$-charge and $M$ and $\tilde{M}$ have charge
$r-2$ and $r+2$, while $(\chi,\tilde{\lambda})$ and
$(\tilde{\chi},\lambda)$ have charge $r-1$ and $r+1$.
They transform under the supercharge $\delta=\delta_{\epsilon}+\delta_{\tilde{\epsilon}}$
as
\begin{align}
\begin{aligned}\delta C & =\frac{i}{2}(\epsilon\chi-\tilde{\epsilon}\tilde{\chi})\ ,\\
\delta\chi & =M\epsilon-\frac{1}{2}(a_{\mu}+iD_{\mu}C)\sigma^{\mu}\tilde{\epsilon}\ ,\\
\delta\tilde{\chi} & =\tilde{M}\tilde{\epsilon}-\frac{1}{2}(a_{\mu}-iD_{\mu}C)\bar{\sigma}^{\mu}\epsilon\ ,\\
\delta M & =\frac{1}{2}\tilde{\epsilon}\bar{\sigma}^{\mu}D_{\mu}\chi+\frac{1}{2}\tilde{\epsilon}\tilde{\lambda}\ ,\\
\delta\tilde{M} & =\frac{1}{2}\epsilon\sigma^{\mu}D_{\mu}\tilde{\chi}+\frac{1}{2}\epsilon\lambda\ ,\\
\delta a_{\mu} & =-\frac{1}{2}D_{\mu}\left(\epsilon\chi+\tilde{\epsilon}\tilde{\chi}\right)-\frac{1}{2}\left(\epsilon\sigma_{\mu}\tilde{\lambda}+\tilde{\epsilon}\bar{\sigma}_{\mu}\lambda\right)\ ,\\
\delta\lambda & =\frac{1}{2}(\sigma^{\mu\nu}F_{\mu\nu}+iD)\epsilon\ ,\\
\delta\tilde{\lambda} & =\frac{1}{2}(\bar{\sigma}^{\mu\nu}F_{\mu\nu}-iD)\tilde{\epsilon}\ ,\\
\delta D & =\frac{i}{2}D_{\mu}\left(\epsilon\sigma^{\mu}\tilde{\lambda}-\tilde{\epsilon}\bar{\sigma}^{\mu}\lambda\right)-\frac{1}{2}V^{\mu}\left(\epsilon\sigma^{\mu}\tilde{\lambda}+\tilde{\epsilon}\bar{\sigma}^{\mu}\lambda\right)+i\frac{r}{8}\left(R-6V_{\mu}V^{\mu}\right)(\epsilon\chi+\tilde{\epsilon}\tilde{\chi})\ ,
\end{aligned}
\label{GeneralSUSYTr}
\end{align}
where the covariant derivatives are defined by 
\begin{align}
\begin{aligned}D_{\mu}C & =(\partial_{\mu}-irA_{\mu})C\ ,\\
D_{\mu}\chi & =\left(\nabla_{\mu}-i\left(r-1\right)A_{\mu}-\frac{i}{2}V_{\mu}\right)\chi\ ,\\
D_{\mu}\tilde{\chi} & =\left(\nabla_{\mu}-i\left(r+1\right)A_{\mu}+\frac{i}{2}V_{\mu}\right)\tilde{\chi}\ .
\end{aligned}
\end{align}
The field strength $F_{\mu\nu}$ is given by
\begin{align}
F_{\mu\nu}=D_{\mu}a_{\nu}-D_{\nu}a_{\mu}\ ,\qquad D_{\mu}a_{\nu}=(\partial_{\mu}-irA_{\mu})a_{\nu}\ .
\end{align}

Given two general multiplets $\CS_{1},\CS_{2}$, we can construct
a new general multiplet $\CS$ whose components are given by \cite{Sohnius:1982fw,Kugo:1982cu}
\begin{align}
\begin{aligned}C & =C_{1}C_{2}\ ,\qquad\chi=C_{1}\chi_{2}+C_{2}\chi_{1}\ ,\qquad\tilde{\chi}=C_{1}\tilde{\chi}_{2}+C_{2}\tilde{\chi}_{1}\ ,\\
M & =C_{1}M_{2}+C_{2}M_{1}-\frac{i}{2}\chi_{1}\chi_{2}\ ,\qquad\tilde{M}=C_{1}\tilde{M}_{2}+C_{2}\tilde{M}_{1}+\frac{i}{2}\tilde{\chi}_{1}\tilde{\chi}_{2}\ ,\\
a_{\mu} & =C_{1}a_{\mu}^{2}+C_{2}a_{\mu}^{1}+\frac{i}{2}(\chi_{1}\sigma_{\mu}\tilde{\chi}_{2}-\tilde{\chi}_{1}\bar{\sigma}_{\mu}\chi_{2})\ ,\\
\lambda & =\left(C_{1}\lambda_{2}-i\tilde{M}_{1}\chi_{2}+\frac{i}{2}(a_{\mu}^{1}+iD_{\mu}C_{1})\sigma^{\mu}\tilde{\chi}_{2}\right)+(1\leftrightarrow2)\ ,\\
\tilde{\lambda} & =\left(C_{1}\tilde{\lambda}_{2}-iM_{1}\tilde{\chi}_{2}-\frac{i}{2}(a_{\mu}^{1}-iD_{\mu}C_{1})\bar{\sigma}^{\mu}\chi_{2}\right)+(1\leftrightarrow2)\ ,\\
D & =C_{1}D_{2}+C_{2}D_{1}+2(M_{1}\tilde{M}_{2}+\tilde{M}_{1}M_{2})-a_{\mu}^{1}a_{2}^{\mu}-D_{\mu}C_{1}D^{\mu}C_{2}\\
 & \qquad-\frac{1}{2}\left(2\chi_{1}\lambda_{2}+2\tilde{\chi}_{1}\tilde{\lambda}_{2}+\chi_{1}\sigma^{\mu}D_{\mu}\tilde{\chi}_{2}+\tilde{\chi}_{1}\bar{\sigma}^{\mu}D_{\mu}\chi_{2}+(1\leftrightarrow2)\right)\\
 & \qquad\quad-iV_{\mu}(\chi_{1}\sigma^{\mu}\tilde{\chi}_{2}-\tilde{\chi}_{1}\bar{\sigma}^{\mu}\chi_{2})\ .
\end{aligned}
\label{MultLaw}
\end{align}

\subsection{Vector multiplet}

A vector multiplet $\CV$ has no $R$-charge. Its embedding in a
general multiplet, in Wess-Zumino gauge, is 
\begin{align}
\CV=\left(0,0,0,0,0,a_{\mu},\lambda,\tilde{\lambda},D\right)\ ,
\end{align}
which transforms under the supersymmetry as 
\begin{align}
\begin{aligned}\delta a_{\mu} & =-\frac{1}{2}(\epsilon\sigma_{\mu}\tilde{\lambda}+\tilde{\epsilon}\bar{\sigma}_{\mu}\lambda)\ ,\\
\delta\lambda & =\frac{1}{2}(\sigma^{\mu\nu}F_{\mu\nu}+iD)\epsilon\ ,\\
\delta\tilde{\lambda} & =\frac{1}{2}(\bar{\sigma}^{\mu\nu}F_{\mu\nu}-iD)\tilde{\epsilon}\ ,\\
\delta D & =\frac{i}{2}(\epsilon\sigma^{\mu}D_{\mu}\tilde{\lambda}-\tilde{\epsilon}\bar{\sigma}^{\mu}D_{\mu}\lambda)\ ,
\end{aligned}
\label{VectorTr}
\end{align}
where 
\begin{align}
\begin{aligned}F_{\mu\nu} & =\partial_{\mu}a_{\nu}-\partial_{\nu}a_{\mu}-i[a_{\mu},a_{\nu}]\ ,\\
D_{\mu}\lambda & =\nabla_{\mu}\lambda-i\left(A_{\mu}-\frac{3}{2}V_{\mu}\right)\lambda-i[a_{\mu},\lambda]\ ,\\
D_{\mu}\tilde{\lambda} & =\nabla_{\mu}\tilde{\lambda}+i\left(A_{\mu}-\frac{3}{2}V_{\mu}\right)\tilde{\lambda}-i[a_{\mu},\tilde{\lambda}]\ .
\end{aligned}
\label{eq:GaugeCov}
\end{align}

\subsection{Chiral multiplet}

A chiral multiplet $\Phi^{i}$ of $R$-charge $r_{i}$ and gauge charge
$q_{i}$ is an irreducible representation whose embedding in a general
multiplet is given by 
\begin{align}
\Phi^{i}=\left(\phi^{i},-i\psi^{i},0,-iF^{i},0,iD_{\mu}\phi^{i},0,0,\frac{r_{i}}{4}\left(R-6V_{\mu}V^{\mu}\right)\phi^{i}\right)\ ,
\end{align}
which transforms as 
\begin{align}
\begin{aligned}\delta\phi^{i} & =\frac{1}{2}\epsilon\psi^{i}\ ,\\
\delta\psi^{i} & =\epsilon F^{i}+\sigma^{\mu}\tilde{\epsilon}D_{\mu}\phi^{i}\ ,\\
\delta F^{i} & =\frac{1}{2}\tilde{\epsilon}\bar{\sigma}^{\mu}D_{\mu}\psi^{i}+iq_{i}\tilde{\epsilon}\tilde{\lambda}\phi^{i}\ ,
\end{aligned}
\label{ChiralTr}
\end{align}
where 
\begin{align}
\begin{aligned}D_{\mu}\phi^{i} & =\left(\partial_{\mu}-ir_{i}A_{\mu}-iq_{i}a_{\mu}\right)\phi^{i}\ ,\\
D_{\mu}\psi^{i} & =\left(\nabla_{\mu}-i\left(r_{i}-1\right)A_{\mu}-\frac{i}{2}V_{\mu}-iq_{i}a_{\mu}\right)\psi^{i}\ .
\end{aligned}
\end{align}

An anti-chiral multiplet $\tilde{\Phi}^{\bar{i}}$ of $R$-charge
$r_{\bar{i}}$ and gauge charge $q_{\bar{i}}$ is embedded in a general
multiplet 
\begin{align}
\tilde{\Phi}^{\bar{i}}=\left(\tilde{\phi}^{\bar{i}},0,i\tilde{\psi}^{\bar{i}},0,i\tilde{F}^{\bar{i}},-iD_{\mu}\tilde{\phi}^{\bar{i}},0,0,-\frac{r_{\bar{i}}}{4}\left(R-6V_{\mu}V^{\mu}\right)\tilde{\phi}^{\bar{i}}\right)\ ,
\end{align}
and transforms as 
\begin{align}
\begin{aligned}\delta\tilde{\phi}^{\bar{i}} & =\frac{1}{2}\tilde{\epsilon}\tilde{\psi}^{\bar{i}}\ ,\\
\delta\tilde{\psi}^{\bar{i}} & =\tilde{\epsilon}\tilde{F}^{\bar{i}}+\bar{\sigma}^{\mu}\epsilon D_{\mu}\tilde{\phi}^{\bar{i}}\ ,\\
\delta\tilde{F}^{\bar{i}} & =\frac{1}{2}\epsilon\sigma^{\mu}D_{\mu}\tilde{\psi}^{\bar{i}}-iq_{\bar{i}}\epsilon\lambda\tilde{\phi}^{\bar{i}}\ ,
\end{aligned}
\label{AntiChiralTr}
\end{align}
where 
\begin{align}
\begin{aligned}D_{\mu}\tilde{\phi}^{\bar{i}} & =\left(\partial_{\mu}-ir_{\bar{i}}A_{\mu}-iq_{\bar{i}}a_{\mu}\right)\tilde{\phi}^{\bar{i}}\ ,\\
D_{\mu}\tilde{\psi}^{\bar{i}} & =\left(\nabla_{\mu}-i\left(r_{\bar{i}}+1\right)A_{\mu}+\frac{i}{2}V_{\mu}-iq_{\bar{i}}a_{\mu}\right)\tilde{\psi}^{\bar{i}}\ .
\end{aligned}
\end{align}

There is another useful (anti-)chiral multiplet constructed from a
vector multiplet, $W_{\alpha}$ ($\tilde{W}_{\alpha}$). They have
$R$-charge $r=1$ and $-1$, respectively, and the components are
given by 
\begin{align}\label{chiralFieldStrength}
\begin{aligned}W_{\alpha} & =\left(\lambda_{\alpha},(\sigma_{\mu\nu})_{\beta\alpha}F^{\mu\nu}+iD\varepsilon_{\beta\alpha},(\sigma^{\mu}D_{\mu}\tilde{\lambda})_{\alpha}\right)\ ,\\
\tilde{W}^{\dot{\alpha}} & =\left(\tilde{\lambda}^{\dot{\alpha}},(\bar{\sigma}_{\mu\nu})^{\dot{\beta}\dot{\alpha}}F^{\mu\nu}-iD\varepsilon^{\dot{\beta}\dot{\alpha}},(\bar{\sigma}^{\mu}D_{\mu}\lambda)^{\dot{\alpha}}\right)\ .
\end{aligned}
\end{align}

\subsection{Real linear multiplet}

A real linear multiplet does not have $R$-charge ($r=0$). Its
embedding in a general multiplet is 
\begin{align}
\CJ=(J,j,\tilde{j},0,0,j_{\mu}+2V_{\mu}J,-\sigma^{\mu}D_{\mu}\tilde{j},-\bar{\sigma}^{\mu}D_{\mu}j,-\nabla_{\mu}\nabla^{\mu}J-2V_{\mu}j^{\mu})\ .
\end{align}
The vector component $j_{\mu}$ is a conserved current 
\begin{align}
\nabla_{\mu}j^{\mu}=0\ .
\end{align}
The transformation law of a linear multiplet is given by 
\begin{align}
\begin{aligned}\delta J & =\frac{i}{2}(\epsilon j-\tilde{\epsilon}\tilde{j})\ ,\\
\delta j & =-\frac{1}{2}\left(j_{\mu}+i\nabla_{\mu}J+2V_{\mu}J\right)\sigma^{\mu}\tilde{\epsilon}\ ,\\
\delta\tilde{j} & =-\frac{1}{2}(j_{\mu}-i\nabla_{\mu}J+2V_{\mu}J)\bar{\sigma}^{\mu}\epsilon\ ,\\
\delta j_{\mu} & =-\nabla^{\nu}(\epsilon\sigma_{\mu\nu}j+\tilde{\epsilon}\bar{\sigma}_{\mu\nu}\tilde{j})\ .
\end{aligned}
\end{align}

\subsection{Supersymmetric Lagrangians}

One can construct an invariant action by integrating the $D$-term of a general multiplet with no $R$-charge 
\begin{align}
\CL_{D}=D-V_{\mu}a^{\mu}\ ,
\end{align}
or the $F$-terms of a chiral multiplet of $r=2$ and an anti-chiral
multiplet of $r=-2$ 
\begin{align}
\CL_{F}=F+\tilde{F}\ .
\end{align}
Invariance follows from the transformation laws \eqref{GeneralSUSYTr} and
\eqref{ChiralTr}.

The $D$-term of a K{\" a}hler potential $K$ whose arguments are chiral and
anti-chiral multiplets $\Phi^{i}$ and $\tilde{\Phi}^{i}$ with
charges $(r_i,q_i=1)$ and $(r_{\bar i},q_{\bar i}=-1)$ 
gives the kinetic Lagrangian of the matter multiplets.
Also, the $F$-terms of superpotentials $W(\Phi)$ and $\tilde{W}(\tilde{\Phi})$
become interactions of the matters. Using the multiplication law \eqref{MultLaw},
the matter Lagrangian is obtained in \cite{Sohnius:1982fw,Cremmer:1982en}
\begin{align}
\begin{aligned}\CL_{\text{matter}} & =-[K(\Phi,\tilde{\Phi})]_{D}-[W(\Phi)]_{F}-[\tilde{W}(\tilde{\Phi})]_{\tilde{F}}\ ,\\
 & =-\left(\frac{1}{2}R-3V_{\mu}^{2}\right)\left(\frac{1}{4}r_{i}K_{i}\phi^{i}-\frac{1}{4}r_{\bar{i}}K_{\bar{i}}\tilde{\phi}^{\bar{i}}\right)+K_{i\bar{j}}(D_{\mu}\phi^{i}D^{\mu}\tilde{\phi}^{\bar{j}}-F^{i}\tilde{F}^{\bar{j}})\\
 & \qquad\quad-iV^{\mu}(K_{i}D_{\mu}\phi^{i}-K_{\bar{i}}D_{\mu}\tilde{\phi}^{\bar{i}})-F^{i}W_{i}-\tilde{F}^{\bar{i}}\tilde{W}_{\bar{i}}-K^{a}D^{a}\\
 & \quad+\frac{1}{2}K_{i\bar{j}}\tilde{\psi}^{\bar{j}}\bar{\sigma}^{\mu}D_{\mu}\psi^{i}+\frac{1}{4}K_{ij\bar{j}}\tilde{F}^{\bar{j}}\psi^{i}\psi^{j}+\frac{1}{4}K_{\overline{ij}j}F^{j}\tilde{\psi}^{\bar{i}}\tilde{\psi}^{\bar{j}}\\
 & \qquad\quad+\frac{1}{4}W_{ij}\psi^{i}\psi^{j}+\frac{1}{4}\tilde{W}_{\overline{ij}}\tilde{\psi}^{\bar{i}}\tilde{\psi}^{\bar{j}}-\frac{1}{16}K_{ij\overline{ij}}\psi^{i}\psi^{j}\tilde{\psi}^{\bar{i}}\tilde{\psi}^{\bar{j}}-i\left(\lambda^{a}K_{i}^{a}\psi^{i} - \tilde{\lambda}^{a}K_{\bar{i}}^{a}\tilde{\psi}^{\bar{i}}\right)\ ,
\end{aligned}
\label{Matter_Lag}
\end{align}
where we denote $K_{i}=\partial_{\phi^{i}}K(\phi)$ and so on, and
defined 
\begin{align}
K^{a}=\tilde{\phi}^{\bar{i}}\, T_{\bar{i}\bar{j}}^{a}\, K_{\bar{j}}=K_{i}\, T_{ij}^{a}\,\phi^{j}\ ,
\end{align}
and $T^{a}$ is a generator of a gauge group. This action agrees
with the Lagrangian derived from the rigid limit of the new minimal
supergravity \cite{Festuccia:2011ws} up to the difference of the
conventions. Here the covariant derivative and the Christoffel symbol
are defined by 
\begin{align}
\begin{aligned}D_{\mu}\psi^{i} & =\left(\nabla_{\mu}-i\left(r_{i}-1\right)A_{\mu}-\frac{i}{2}V_{\mu}\right)\psi^{i}-iT_{ij}^{a}a_{\mu}^{a}\psi^{j}+\Gamma_{jk}^{i}\psi^{j}D_{\mu}\phi^{k}\ ,\\
\Gamma_{jk}^{i} & =K^{i\bar{i}}K_{\bar{i}jk}\ ,\\
D_{\mu}\phi^{i} & =\left(\partial_{\mu}-ir_{i}A_{\mu}\right)\phi^{i}-iT_{ij}^{a}a_{\mu}^{a}\phi^{j}\ .
\end{aligned}
\end{align}

Similarly, the Lagrangian of the gauge sector is given by using the
field strength chiral multiplet \eqref{chiralFieldStrength} as \cite{Cremmer:1982en}
\begin{align}
\begin{aligned}\CL_{\text{gauge}} & =\frac{1}{2}[f_{AB}(\Phi)W^{A}W^{B}]_{F}+\frac{1}{2}[\tilde{f}_{AB}(\tilde{\Phi})\tilde{W}^{A}\tilde{W}^{B}]_{\tilde{F}}\ ,\\
 & =\text{Tr}\left[\frac{1}{4}(f_{AB}+\tilde{f}_{AB})F_{\mu\nu}^{A}F^{B\,\mu\nu}-\frac{1}{8}(f_{AB}-\tilde{f}_{AB})\varepsilon^{\mu\nu\rho\kappa}F_{\mu\nu}^{A}F_{\rho\kappa}^{B}\right.\\
 & \qquad\quad+f_{AB}\lambda^{A}\sigma^{\mu}D_{\mu}\tilde{\lambda}^{B}+\tilde{f}_{AB}\tilde{\lambda}^{A}\bar{\sigma}^{\mu}D_{\mu}\lambda^{B}-\frac{1}{2}(f_{AB}+\tilde{f}_{AB})D^{A}D^{B}\\
 & \qquad\quad+\frac{1}{2}f_{AB,i}\left(F^{i}\lambda^{A}\lambda^{B}-iD^{A}\psi^{i}\lambda^{B}+F_{\mu\nu}^{A}\psi^{i}\sigma^{\mu\nu}\lambda^{B}\right)\\
 & \qquad\quad+\frac{1}{2}\tilde{f}_{AB,\bar{i}}\left(\tilde{F}^{\bar{i}}\tilde{\lambda}^{A}\tilde{\lambda}^{B}+iD^{A}\tilde{\psi}^{\bar{i}}\tilde{\lambda}^{B}+F_{\mu\nu}^{A}\tilde{\psi}^{\bar{i}}\bar{\sigma}^{\mu\nu}\tilde{\lambda}^{B}\right)\\
 & \qquad\quad\left.-\frac{1}{8}f_{AB,ij}\lambda^{A}\lambda^{B}\psi^{i}\psi^{j}-\frac{1}{8}\tilde{f}_{AB,\overline{ij}}\tilde{\lambda}^{A}\tilde{\lambda}^{B}\tilde{\psi}^{\bar{i}}\tilde{\psi}^{\bar{j}}\right]\ ,
\end{aligned}
\label{Gauge_Lag}
\end{align}
where $f_{AB}(\Phi)$ and $\tilde{f}_{AB}(\tilde{\Phi})$ are functions
of the matter fields, and $A,B$ label the types of gauge groups. 
The field strength and the covariant derivatives are defined by \eqref{eq:GaugeCov}.

\section{Localization}\label{ss:Localization}
So far we have described the $\CN =1$ supersymmetry multiplets and Lagrangians on Hermitian manifolds that admit at least one supercharge by taking the rigid limit of the new minimal supergravity.
The most general supersymemtric action, which we denote $S$, is given by the spacetime integral of the Lagrangians \eqref{Matter_Lag} and \eqref{Gauge_Lag}. 
To compute the partition function on a Hermitian manifold $M$ by localization, we add a $\delta$-exact term $\delta V$ to the action and compute the deformed partition function
\begin{align}
Z(t) = \int \CD \phi \, e^{-S - t \delta V} \ .
\end{align}
Since $Z(t)$ does not depend on the parameter $t$, we let $t$ be large while choosing a positive semi-definite $\delta V$. The integral localizes to the field configurations for which $\delta V$ vanishes. 
We will construct such a localizing term below and perform the localization calculation around the fixed point with the equivariant index theorem in Section \ref{ss:IndexTheorem}.

\subsection{Localization and fixed points}

We consider manifolds with two supercharges $\epsilon$ and $\tilde{\epsilon}$
of opposite $R$-charges. We will use a linear combination of the
two supercharges $\delta=\delta_{\epsilon}+\delta_{\tilde{\epsilon}}$
which is not nilpotent, but satisfies 
\begin{align}
\{\delta,\delta\}=\delta_{K}\ .
\end{align}
We will find a localizing term of the form $\delta V$.

\subsubsection{Localizing terms}

To find a localizing term, we use the normalized complex conjugates of the Killing spinors \eqref{ZetaSpinor}.
The $R$-charges for $\zeta$ and $\tilde{\zeta}$ are $-1$ and $+1$
so as to be consistent with the normalization conditions. They are
invariant under the $R$-covariant Lie derivative along $K$, 
\begin{align}
\CL_{K}^{A}\,\zeta=\CL_{K}^{A}\,\tilde{\zeta}=0\ .\label{LieZeta}
\end{align}
This property is useful to construct localizing terms as follows.
Let $\Psi$ and $\tilde{\Psi}$ be fermionic functions of fields with
$R$-charge $+1$ and $-1$. Consider a Lagrangian density
$v=(\zeta\Psi+\tilde{\zeta}\tilde{\Psi})$ whose $R$-charge vanishes.
Then the spatial integral of $\delta v$ will be a localizing term
because it is $\delta$-closed up to a total derivative 
\begin{align}
\delta^{2}v=\frac{1}{2}\left(\zeta\,\delta_{K}\Psi+\tilde{\zeta}\,\delta_{K}\tilde{\Psi}\right)=\frac{1}{2}\delta_{K} v=(\text{total derivative})\ .
\end{align}
The last equality follows from the fact that $v$ is a scalar function
of zero $R$-charge and $K$ is a Killing vector.

For the gauge sector, we choose the fermionic functions $(\Psi,\tilde{\Psi})$
to be 
\begin{align}
\Psi_{\text{gauge}}=\frac{1}{2}(-F_{\mu\nu}\sigma^{\mu\nu}+iD)\lambda\ ,\qquad\tilde{\Psi}_{\text{gauge}}=\frac{1}{2}(-F_{\mu\nu}\bar{\sigma}^{\mu\nu}-iD)\tilde{\lambda}\ ,
\end{align}
and obtain the localizing term 
\begin{align}
\begin{aligned}\CL_{\text{gauge}}^{(\text{loc})}=\frac{1}{2}F_{\mu\nu}F^{\mu\nu}+\lambda\sigma^{\mu}D_{\mu}\tilde{\lambda}+\tilde{\lambda}\bar{\sigma}^{\mu}D_{\mu}\lambda-D^{2}\ .\end{aligned}
\label{Loc_Gauge}
\end{align}
A positive definite contour is achieved by taking $a_{\mu}$ real
and rotating 
\[
D\rightarrow-iD\ .
\]
We will implicitly substitute $-iD$ for $D$ in all later equations.
The field configurations to which the path integral localizes are those which satisfy
\begin{align}
F_{\mu\nu}=0\ ,\qquad D=0\ .\label{Gauge_Loc}
\end{align}
The localizing term \eqref{Loc_Gauge} is nothing but the Lagrangian
of the gauge sector \eqref{Gauge_Lag} with $f_{AB}=\tilde{f}_{AB}=\delta_{AB}$.
The first condition leads to a flat connection of the gauge field
which will be described in detail in Section \ref{sub:The-bosonic-moduli-space}.

We can fix the gauge freedom by imposing the covariant gauge $\nabla^{\mu}a_{\mu}=0$
\begin{align}
\CL_{\text{g.f.}}=\bar{c}\,\nabla_{\mu}D^{\mu}c+b\nabla^{\mu}a_{\mu}\ ,
\end{align}
where $c$ and $\bar{c}$ are ghost fields and $b$ is a Lagrange
multiplier. As is explained in \cite{Pestun:2007rz,Kapustin:2009kz},
the gauge fixing term does not change the locus of fixed points where
the bosonic part of \eqref{Loc_Gauge} vanishes.

Next, we move onto the matter sector coupled to the gauge field. We
consider a localizing term for the matter sector with a chiral multiplet
$\Phi=(\phi,\psi,F)$ of $R$-charge $r$ and gauge charge $q$ and
an anti-chiral multiplet $\tilde{\Phi}=(\tilde{\phi},\tilde{\psi},\tilde{F})$
of $R$-charge $-r$ and gauge charge $-q$. We choose the fermionic
functions $(\Psi,\tilde{\Psi})$ to be 
\begin{align}
\begin{aligned}\Psi_{\text{matter}}=-\frac{1}{2}(\phi\sigma^{\mu}D_{\mu}\tilde{\psi}+\psi\tilde{F}-2i q\tilde{\phi}\lambda\phi)\ ,\qquad\tilde{\Psi}_{\text{matter}}=-\frac{1}{2}(\tilde{\phi}\bar{\sigma}^{\mu}D_{\mu}\psi+\tilde{\psi}F+2iq\tilde{\phi}\tilde{\lambda}\phi)\ ,\end{aligned}
\end{align}
which yields the localizing term 
\begin{align}
\begin{aligned}\CL_{\text{matter}}^{(\text{loc})} & =D_{\mu}\tilde{\phi}D^{\mu}\phi-iV^{\mu}(\tilde{\phi}D_{\mu}\phi-\phi D_{\mu}\tilde{\phi})+ iq\tilde{\phi}D\phi-\frac{r}{4}\left(R-6V_{\mu}V^{\mu}\right)\phi\tilde{\phi}\\
 & \quad+\frac{1}{2}\tilde{\psi}\bar{\sigma}^{\mu}D_{\mu}\psi-iq\left(\tilde{\phi}\lambda\psi- \tilde{\psi}\tilde{\lambda}\phi\right)-F\tilde{F}\ ,
\end{aligned}
\label{Loc_Mat}
\end{align}
where we used the integrability conditions \eqref{IntegCond} and
\eqref{IntegCondBar} and removed a total derivative term. This agrees
with the matter Lagrangian \eqref{Matter_Lag} with a canonical K{\" a}hler
potential $K=\tilde{\Phi}\Phi$ and without superpotentials. Every
covariant derivative is also covariant with respect to the gauge field
for the matter fields with gauge charges.

The bosonic part of the $\delta$-exact Lagrangian is positive definite
if we choose the contour of the path integral for $\tilde{\phi}$
and $\tilde{F}$ to be 
\begin{align}
\tilde{\phi}=\phi^{\dagger}\ ,\qquad\tilde{F}=-F^{\dagger}\ .\label{IntegContour}
\end{align}
Then the field configuration of the matter sector localizes to 
\begin{align}
\phi=0\ ,\qquad F=0\ .\label{Mat_Locus}
\end{align}
where we used the condition \eqref{Gauge_Locus_Single} for the gauge
sector.

The classical contributions from the gauge and matter Lagrangians
\eqref{Gauge_Lag} and \eqref{Matter_Lag} vanish on the zero loci
\eqref{Gauge_Loc} and \eqref{Mat_Locus}. A classical contribution from a Fayet-Illiopolous term for an abelian gauge multiplet is discussed in \ref{classical_contributions}.

\subsubsection{Cohomological derivation}
Our constructions of the localizing terms are somewhat heuristic and one may wonder if there are other choices.
Here we present an alternative derivation of the fixed points based on the cohomological forms of the supersymmetry transformations.
This approach is taken by \cite{Kallen:2011ny,Ohta:2012ev} for $\CN =2$ Chern-Simons-matter theories on Seifert three-manifolds.

For the vector multiplet \eqref{VectorTr}, we introduce new variables
$(\Lambda_{\mu}^{+},\Lambda_{\mu}^{-})$ defined by 
\begin{align}
\begin{aligned}\Lambda_{\mu}^{+} & \equiv-\frac{1}{2}\epsilon\sigma_{\mu}\tilde{\lambda}\ ,\qquad\tilde{\lambda}=\Lambda_{\mu}^{+}(\zeta\sigma^{\mu})\ ,\\
\ \Lambda_{\mu}^{-} & \equiv-\frac{1}{2}\tilde{\epsilon}\bar{\sigma}_{\mu}\lambda\ ,\qquad\lambda=\Lambda_{\mu}^{-}(\sigma^{\mu}\tilde{\zeta})\ ,
\end{aligned}
\end{align}
which satisfy the supersymmetry transformation 
\begin{align}
\begin{aligned}
\delta_{\epsilon} a_{\mu} & =\Lambda_{\mu}^{+}\ ,\qquad & \delta_{\tilde \epsilon}a_{\mu} & =\Lambda_{\mu}^{-}\ ,\\
\delta_{\epsilon}\Lambda_{\mu}^{+} & =0\ ,\qquad & \delta_{\tilde \epsilon}\Lambda_{\mu}^{+} & =\frac{1}{4}K^{\nu}F_{\nu\mu}^{+}+\frac{1}{4}K_{\mu}D\ ,\\
\delta_{\epsilon}\Lambda_{\mu}^{-} & =\frac{1}{4}K^{\nu}F_{\nu\mu}^{-}-\frac{1}{4}K_{\mu}D\ ,\qquad & \delta_{\tilde \epsilon}\Lambda_{\mu}^{-} & =0\ ,\\
\delta_{\epsilon} D & =-D^{\mu}\Lambda_{\mu}^{+}\ ,\qquad & \delta_{\tilde \epsilon}D & =D^{\mu}\Lambda_{\mu}^{-}\ ,
\end{aligned}
\end{align}
where $F_{\mu\nu}^{\pm}$ is the (anti)self-dual part of the field
strength 
\begin{align}
F_{\mu\nu}^{\pm}\equiv F_{\mu\nu}\pm\frac{1}{2}\varepsilon_{\mu\nu\rho\sigma}F^{\rho\sigma}\ .
\end{align}

For the chiral multiplets, we introduce the bosonic variables $(\psi_{+},\psi_{-})$
and $(\tilde{\psi}_{+},\tilde{\psi}_{-})$ made of the fermions satisfying
\begin{align}
\begin{aligned}\psi & =\epsilon\psi_{+}-\zeta\psi_{-}\ ,\qquad\psi_{+}=\zeta\psi\ ,\qquad\psi_{-}=\epsilon\psi\ ,\\
\tilde{\psi} & =\tilde{\epsilon}\tilde{\psi}_{+}-\tilde{\zeta}\tilde{\psi}_{-}\ ,\qquad\tilde{\psi}_{+}=\tilde{\zeta}\tilde{\psi}\ ,\qquad\tilde{\psi}_{-}=\tilde{\epsilon}\tilde{\psi}\ .
\end{aligned}
\end{align}
We rewrite the supersymmetry transformations of the chiral multiplet
\eqref{ChiralTr} of gauge charge $q$ in the cohomological forms as
\begin{align}
\begin{aligned}\delta_{\epsilon}\phi & =\psi_{-}\ ,\qquad & \delta_{\tilde \epsilon}\phi & =0\ ,\\
\delta_{\epsilon}\psi_{-} & =0\ ,\qquad & \delta_{\tilde \epsilon}\psi_{-} & =\CL_{K}^{A}\phi\ ,\\
\delta_{\epsilon}\psi_{+} & =F\ ,\qquad & \delta_{\tilde \epsilon}\psi_{+} & =\CL_{\bar{Y}}\phi\ ,\\
\delta_{\epsilon} F & =0\ ,\qquad & \delta_{\tilde \epsilon}F & =\frac{1}{2}\CL_{K}^{A}\psi_{+}-\frac{1}{2}\CL_{\bar{Y}}\psi_{-}-iq\bar{Y}^{\mu}\Lambda_{\mu}^{+}\phi\ ,
\end{aligned}
\end{align}
where the vectors $Y^{\mu},\bar{Y}^{\mu}$ are defined in \eqref{DefY}. Similarly the
anti-chiral multiplet \eqref{AntiChiralTr} of gauge charge $-q$
with the new variables transforms as 
\begin{align}
\begin{aligned}\delta_{\epsilon}\tilde{\phi} & =0\ ,\qquad & \delta_{\tilde \epsilon}\tilde{\phi} & =\tilde{\psi}_{-}\ ,\\
\delta_{\epsilon}\tilde{\psi}_{-} & =\CL_{K}^{A}\tilde{\phi}\ ,\qquad & \delta_{\tilde \epsilon}\tilde{\psi}_{-} & =0\ ,\\
\delta_{\epsilon}\tilde{\psi}_{+} & =\CL_{Y}\tilde{\phi}\ ,\qquad & \delta_{\tilde \epsilon}\tilde{\psi}_{+} & =\tilde{F}\ ,\\
\delta_{\epsilon}\tilde{F} & =\frac{1}{2}\CL_{K}^{A}\tilde{\psi}_{+}-\frac{1}{2}\CL_{Y}\tilde{\psi}_{-}+iqY^{\mu}\Lambda_{\mu}^{-}\tilde{\phi}\ ,\qquad & \delta_{\tilde \epsilon}\tilde{F} & =0\ .
\end{aligned}
\end{align}

The localizing terms with respect to the supercharge $\delta=\delta_{\epsilon}+\delta_{\tilde \epsilon}$
are given in this representation as 
\begin{align}\label{LocalizingCohom}
\begin{aligned}v_{\text{gauge}} & =(\delta_{\tilde \epsilon}\Lambda_{\mu}^{+})^{\dagger}\Lambda^{+\mu}+(\delta_{\epsilon}\Lambda_{\mu}^{-})^{\dagger}\Lambda^{-\mu}\ ,\\
v_{\text{matter}} & =(\delta_{\tilde\epsilon}\psi_{+})^{\dagger}\psi_{+}+(\delta_{\tilde\epsilon}\tilde\psi_{+})^{\dagger}\tilde\psi_{+}+4(\delta_{\tilde\epsilon}\psi_{-})^{\dagger}\psi_{-}\ ,
\end{aligned}
\end{align}
leading to the same fixed loci on the integration contour \eqref{IntegContour} as in the previous subsection
\begin{align}
\begin{aligned}F_{\mu\nu} & =0\ ,\qquad & D & =0\ ,\\
\phi &=  0\ ,\qquad & F & =0\ .
\end{aligned}
\end{align}
Note that the saddle points 
\[
D=\frac{1}{2}\varepsilon_{\mu\nu\rho\sigma}K^{\mu}\bar{K}^{\nu}F^{\rho\sigma}\ ,
\]
are off the contour of integration.

\subsubsection{Single supercharge case}

In general, a Hermitian manifold admits a single supercharge $\epsilon$ \cite{Dumitrescu:2012ha}.
The supersymmetry transformation $\delta_{\epsilon}$ is nilpotent
$\delta_{\epsilon}^{2}=0$. It is straightforward to construct a $\delta_\epsilon$-exact
term for the gauge sector 
\begin{align}
\begin{aligned}\CL_{\text{gauge}}^\text{(loc)} & =\delta_{\epsilon}\left(\zeta(-F_{\mu\nu}\sigma^{\mu\nu}+iD)\lambda\right)\ ,\\
 & =(F_{\mu\nu}^{-})^{2}+2\lambda\sigma^{\mu}D_{\mu}\tilde{\lambda}+D^{2}\ .
\end{aligned}
\label{Loc_Gauge-1}
\end{align}
A similar $\delta$-exact term can be derived with a supercharge $\tilde{\epsilon}$
of opposite $R$-charge. In either way, the gauge sector localizes
to a (anti-)self-dual field strength configuration 
\begin{align}
F_{\mu\nu}^{\pm}=0\ ,\qquad D=0\ .\label{Gauge_Locus_Single}
\end{align}
The same result can be derived from the cohomological forms \eqref{LocalizingCohom} by setting either $\epsilon$ or $\tilde\epsilon$ to zero.

The matter sector localizes to the same configuration $\phi=F=0$ as the case with two supercharges of opposite $R$-charge because only one supercharge $\tilde \epsilon$ appears in the cohomological localization term \eqref{LocalizingCohom}.
If a supercharge $\epsilon$ exists on a Hermitian manifold, the following localization term yields the same fixed points
\begin{align}
v_{\text{matter}} & =(\delta_{\epsilon}\tilde{\psi}_{+})^{\dagger}\tilde{\psi}_{+}+(\delta_{\epsilon}\psi_{+})^{\dagger}\psi_{+}+4(\delta_{\epsilon}\tilde{\psi}_{-})^{\dagger}\tilde{\psi}_{-} \ .
\end{align}

\subsection{Index theorem ingredients}

The quantum fluctuations around the zero locus of the localizing terms
in the previous section contribute to the partition function. We will
use the equivariant index theorem for transversally elliptic operators
to compute the one-loop determinant \cite{Pestun:2007rz,Gomis:2011pf}.
To this end, we rewrite the localizing terms as 
\begin{align}
v=(\hat{\varphi}_{o},\varphi_{o})\left(\begin{array}{cc}
D_{\hat{o}e} & D_{\hat{o}\hat{e}}\\
D_{oe} & D_{o\hat{e}}
\end{array}\right)\left(\begin{array}{c}
\varphi_{e}\\
\hat{\varphi}_{e}
\end{array}\right)\ ,
\end{align}
where $\varphi_{e,o}$ are bosonic and fermionic fields, respectively,
and $\hat{\varphi}_{e,o}$ are their $\delta$ variations 
\begin{align}
\begin{aligned}\delta\varphi_{e,o} & =\hat{\varphi}_{o,e}\ ,\\
\delta\hat{\varphi}_{o,e} & =\CR\cdot\varphi_{e,o}\ .
\end{aligned}
\end{align}
$D_{\hat{o}e,\hat{o}\hat{e},oe,o\hat{e}}$ are differential operators
and $\CR$ is a symmetry of the theory. The one-loop determinant is
given by the following expression \cite{Pestun:2007rz}: 
\begin{align}\label{OneLoopPF}
Z_{\text{one-loop}}=\frac{\det_{\,\text{Coker}D_{oe}}\CR}{\det_{\,\text{Ker}D_{oe}}\CR}\ .
\end{align}

For the matter sector, we identify
\[
D_{oe}=\mathcal{L}_{\bar Y} \ ,
\]
acting on the sections $\phi$, and
\[
\mathcal{R}=\delta_{K}\ .
\]
In complex coordinates, the relevant vectors are given by \eqref{KillingVectors}, hence 
\[
D_{oe}=\frac{s}{c}\left(\partial_{\bar{z}}-\bar{h}\left(z,\bar{z}\right)\partial_{\bar{w}}-ir A_{\bar{z}}+ir \bar{h}\left(z,\bar{z}\right)A_{\bar{w}}-iq_{f}a_{\bar{z}}+iq_{f}\bar{h}\left(z,\bar{z}\right)a_{\bar{w}}\right) \ .
\]
We check that this commutes with the symmetry generated by

\[
\left[\delta_{\epsilon},\delta_{\tilde{\epsilon}}\right]=\delta_{K}=\mathcal{L}_{K}-ir K^{\mu}A_{\mu}-iq_{f}K^{\mu}a_{\mu} \ ,
\]
indeed
\[
\left[\delta_{K},D_{oe}\right]=ir F_{\bar{z}\bar{w}}^{A}+iq_{f}F_{\bar{z}\bar{w}}^{a}=0\ .
\]
Note that this remains true if we allow a flux $F_{z\bar{z}}^{a}$. 

The leading symbol of $D_{oe}$ is just $\bar{Y}^{\mu}$. 
The equations \eqref{KYrelation} imply that this is non-zero on the subspace spanned by the (non-vanishing) vectors
\[
\text{Re}\left(Y^{\mu}\right)\ , \qquad \text{Im}\left(Y^{\mu}\right) \ ,
\]
which form a basis for the subspace orthogonal to $K,\bar{K}$. Hence
$D_{oe}$ is transversally elliptic. Note that this subspace need
not be two-dimensional. We will show in Section \ref{ss:IndexTheorem} that the correct fluctuation determinant is recovered by viewing $D_{oe}$ as the pullback of the \emph{Dirac} operator on the base manifold $\Sigma$, or the $\bar{\partial}$ operator twisted by the square root of the canonical bundle, acting on sections with $R$-charge $r-1$. This can be argued for by considering the form of the vector $\bar{Y}^\mu$ in local coordinates. However, we do not have a completely satisfactory derivation of this fact.  

For the gauge sector, 
\[
D_{oe}= \iota_Y \iota_{\bar{Y}} d \ ,
\] 
is considered as a differential operator acting on the connection $a_\mu$, where $\iota$ is the interior product and $d$ the exterior derivative.  The commutator of this operator with $\delta_K$ contains the same types of terms as the matter sector operator above and therefore vanishes on the moduli space. To prove transversal ellipticity one must consider the combined supersymmetry and BRST complex. We describe this complex in Section \ref{ss:IndexTheorem}. We will also find that to recover the correct fluctuation determinant, this operator should be identified with the pullback of the exterior derivative on $\Sigma$.

\subsubsection{The bosonic moduli space}\label{sub:The-bosonic-moduli-space}

We have shown that the bosonic part of the localization locus is the
moduli space of flat $G$-connections on $M$. The partition function
on $M$ contains an integral over this space, which may have many
connected components. Background deformations associated with flavor
symmetries are just flat background gauge fields.

Flat connections are specified by holonomies. The formula for the
one-loop determinants given by the equivariant index theorem implies
that we must determine how such holonomies affect the operators $D_{oe}$
and $\delta_{K}$. Since the equivariant index depends only on discrete
parameters specifying the spaces (bundles) which $D_{oe}$ maps, we
should find out how to associate a bundle on $\Sigma$ to every holonomy.
The operator $\delta_{K}$ depends on continuous data related to holomorphic
moduli. nontrivial moduli arise when there is a holonomy in one or
more of the directions corresponding to the circle actions of the
Killing vector $K$. What remains is to determine the allowed bundles
and continuous moduli. The manifold $M$ admits a much larger space
of flat connections then that which $D_{oe}$ and $\delta_{K}$ can
``measure''. An element of the moduli space of flat connections
which deforms neither $D_{oe}$ nor $\delta_{K}$ (and which does
not give a classical contribution) will still contribute to the normalization
of parts of the partition function on $M$.

The moduli space of flat connections on $M$ is given by 
\begin{align}
\mathcal{M}_{G}^{0}\left(M\right)=\text{Hom}\left(\pi_{1}\left(M\right),G\right)/G\ ,\label{eq:moduli_space_of_flat_connections}
\end{align}
where the holonomy associated to a generator $a\in\pi_{1}\left(M\right)$
is given by 
\[
\sigma\left(a\right)\in G\ ,\qquad\sigma\in\text{Hom}\left(\pi_{1}\left(M\right),G\right)\ ,
\]
and the quotient is taken with $G$ acting on all $\sigma\left(\bullet\right)$
by conjugation. The first step in characterizing this space is to
compute the fundamental group of $M$. Note that 
\[
M\simeq M_{3}\times S^{1}\quad\Rightarrow\quad\pi_{1}\left(M\right)=\pi_{1}\left(M_{3}\right)\times\mathbb{Z}\ .
\]
$M_{3}$ is a circle bundle of degree $d$ over $\Sigma$ with the Euler characteristic
\[
\chi\left(\Sigma\right)=2-2g\ .
\]
For $g=0$ it is a lens space $L\left(d,1\right)$ with 
\begin{align}
\pi_{1}\left(M_{3}\right)=\mathbb{Z}_{d}\ .\label{eq:lens_space_fundamental_group}
\end{align}
For $g\ge1$ we have 
\[
\pi_{1}\left(S^{1}\right)=\mathbb{Z}\ ,\qquad\pi_{2}\left(\Sigma\right)=1\ ,
\]
and hence the following short exact sequence holds
\begin{align*}
1\rightarrow\mathbb{Z}\rightarrow\pi_{1}\left(M_{3}\right)\rightarrow\pi_{1}\left(\Sigma\right)\rightarrow1\ .
\end{align*}
There is an explicit presentation for $\pi_{1}\left(M_{3}\right)$
with generators $a_{i},b_{i},h$ with $i\in1,\ldots,g$ and relations
\cite{orlik1972seifert} 
\[
\left[a_{i},h\right]=\left[b_{i},h\right]=1\ ,\qquad\prod_{i=1}^{g}\left[a_{i},b_{i}\right]=h^{d}\ ,
\]
(all other commutators vanish) which reduces to \eqref{eq:lens_space_fundamental_group}
when $g=0$ and whose abelianization is 
\[
H_{1}\left(M_{3},\mathbb{Z}\right)=\mathbb{Z}_{d}\times\mathbb{Z}^{2g}\ .
\]
The fundamental group of $M$ can therefore be described by generators
\[
a_{i},b_{i},h,x,\qquad i\in1,\ldots,g\ ,
\]
and relations 
\begin{align}
\left[a_{i},h\right]=\left[b_{i},h\right]=\left[a_{i},x\right]=\left[b_{i},x\right]=\left[x,h\right]=1\ , \qquad\prod_{i=1}^{g}\left[a_{i},b_{i}\right]=h^{d}\ .\label{eq:fundamental_group_of_M}
\end{align}
For $g=0$, this implies $h^{d}=1$. The relevant space of holonomies
in this case is described in \cite{Razamat:2013opa}.

The above description of $\pi_{1}\left(M_{3}\right)$ is related to
the central extension of $\pi_{1}\left(\Sigma\right)$ considered
by Atiyah and Bott in \cite{atiyah1983yang} in relation to \emph{Yang-Mills}
connections on $\Sigma$. The group element $h^{d}$ plays the role
of $\Gamma_{\mathbb{R}}$. Following \cite{atiyah1983yang}, we will
characterize the set of solutions to \eqref{eq:moduli_space_of_flat_connections}
given the relations \eqref{eq:fundamental_group_of_M} for the case
$G=U\left(N\right)$.
\footnote{See also \cite{Caporaso:2006kk} section 6.2. For an example involving
$G$ with finite, but non trivial, $\pi_{1}\left(G\right)$ see \cite{Razamat:2013opa}.
} 

The space \eqref{eq:moduli_space_of_flat_connections} for $G=U\left(N\right)$
is the space of equivalence classes of unitary representations of $\pi_{1}\left(M\right)$.
Any such representation is the direct sum of irreducible representations,
and within each summand $\sigma\left(h\right)$ and $\sigma\left(x\right)$
are scalar matrices by Schur's lemma. Since there are no further constraints
on $\sigma\left(x\right)$ it may be any such unitary matrix whose
eigenvalue we denote
\[
\exp\left(2\pi ix_{a}\right)\ .
\]
 The possibilities for $\sigma\left(h\right)$ are more restricted
and, in fact, discrete.

Given that the generators $x$ and $h$ commute, we can use $G$ to
simultaneously diagonalize the associated holonomies. Denote by $\lambda_{a}$
the diagonal entries of the matrix $\sigma\left(h\right)$. We define
\[
\lambda_{a}=\exp\left(2\pi ih_{a}\right)\ .
\]
Consider an $N$-dimensional unitary representation of $\pi_{1}\left(M\right)$,
$\mathfrak{R}$, whose decomposition contains $p$ irreducible representations
$\mathfrak{R}_{j}$ of size $N_{j}$ 
\begin{align}\label{eq:symmetry_breaking_partitions_sum}
\begin{aligned}
\sum_{j=1}^{p}N_{j}&=N\ , \\
\mathfrak{R}\simeq\mathfrak{R}_{1}\oplus\mathfrak{R}_{2}&\oplus\cdots\oplus\mathfrak{R}_{p}\ .
\end{aligned}
\end{align}
This induces a symmetry breaking pattern 
\begin{align}
U\left(N\right)\rightarrow U\left(N_{1}\right)\times U\left(N_{2}\right)\times\cdots\times U\left(N_{p}\right)\ .\label{eq:symmetry_breaking_pattern}
\end{align}
Consider the restriction to a particular factor in \eqref{eq:symmetry_breaking_pattern}.
Taking the determinants on both sides of 
\[
\prod_{i=1}^{g}\left[a_{i},b_{i}\right]=h^{d}\ ,
\]
we get the condition 
\[
dN_{j}h_{j\left(a\right)}\in\mathbb{Z}\ ,
\]
where we have introduced the notation $j\left(a\right)$ for the $a$'th
eigenvalue which lies in the $j$'th representation $\mathfrak{R}_{j}$.
The full set of solutions is 
\[
h_{j\left(a\right)}=\frac{m_{j}}{dN_{j}}\ ,\qquad m_{j}\in0, \ldots, dN_{j}-1\ .
\]
 We denote by $\mathcal{M}_{N_{j},m_{j}}^{g}$ the space of \emph{irreducible}
representations satisfying
\[
\prod_{i=1}^{g}\left[a_{i},b_{i}\right]=e^{2\pi i\frac{m_{j}}{N_{j}}}\mathbbm{1}_{N_{j}}\ .
\]
Such representations exist for all $N_{j},m_{j}$ when $g\ge2$, and
for $m_{j}=0,N_{j}=1$ or $m_{j}\ne0$ and $\text{gcd}\left(N_{j},m_{j}\right)=1$
for $g=1$ \cite{atiyah1983yang}.  Some of the representations $\mathfrak{R}_{j}$
may coincide. When the set of representations is discrete, we denote
by $n_{l}$ the multiplicity of the $l$'th representation.  

To understand the interpretation of the solutions above we appeal
to the analysis of Yang-Mills connections on $\Sigma$ given in \cite{atiyah1983yang}.
Translating the data, a particular $\sigma\left(h\right)$ above corresponds
to a homomorphism of the central extension of $\pi_{1}\left(\Sigma\right)$
with a symmetry breaking pattern \eqref{eq:symmetry_breaking_pattern}.
The image of the additional generator $J$ (in the notation of \cite{atiyah1983yang})
\begin{align}
\prod_{i=1}^{g}\left[a_{i},b_{i}\right]=J\ ,\label{eq:yang_mills_central_extension}
\end{align}
in each block is 
\[
\sigma\left(J\right)=e^{2\pi i\frac{m_{j}}{N_{j}}}\mathbbm{1}_{N_{j}}\ .
\]
According to \cite{atiyah1983yang}, the set of unitary representations
of \eqref{eq:yang_mills_central_extension} is isomorphic to the space
of unitary Yang-Mills connections on $\Sigma$. Such connections are
actually $H$-connections 
\[
H\equiv U\left(N_{1}\right)\times U\left(N_{2}\right)\times\cdots\times U\left(N_{p}\right)\ .
\]
Elements of this space are flat $H$-connections twisted by constant
curvature line bundles with first Chern classes 
\begin{align}
c_{1}\left(U\left(N_{j}\right)\right)=m_{j}\ .\label{eq:base_line_bundle_Chern_classes}
\end{align}
The pullback of such a connection on $\Sigma$, augmented with a holonomy
$\sigma\left(x\right)$, is our desired flat connection on $M$. Note
that only the overall Chern class 
\[
c_{1}\left(U\left(N\right)\right)=\sum_{j}m_{j}\text{ mod }d\ ,
\]
is a bundle invariant on $M$. This class resides in the torsion part
of $H^{2}\left(M,\mathbb{Z}\right)$.

Having described the moduli space, we now consider how a set of holonomies
associated to generators of $\pi_{1}\left(M\right)$ deforms the operator
$\delta_{K}$. Recall that $\delta_{K}$ includes a term ($a_{\mu}$
is the $G$-connection, not the generator) 
\[
a_{w}=K^{\mu}a_{\mu}\ .
\]
Since $\delta_{K}$ is supposed to be a torus action, we should expand
the field on which it acts in eigenspaces using the weights of the
$G$-representation.\footnote{In general, such a decomposition need not correspond to the decomposition
in terms of weights. This is because the holonomies for a group which
is not simply-connected may commute without belonging to the same
Cartan torus. See \cite{Razamat:2013opa} for examples. This does
to apply to $G=U\left(N\right)$.
}
We will use this decomposition to diagonalize the action of the commuting
holonomies associated to $h$ and $x$. For $g\ge2$, there is no isometry
action on the base and the holonomies associated with $a_{i},b_{i}$
do not deform $\delta_{K}$, while for $g=0$, $x,h$ are the only
holonomies. In the notation of Section \ref{ss:IndexTheorem},
\[\label{FlavorHolonomy_g>1}
\rho\left(a_{w}\right)=\rho^{a}x_{a}+\tau\rho^{a}h_{a}\ .
\]
For $g=1$, we have the possibility of having both an isometry action
and nontrivial holonomies for a flat connection on $\Sigma$. We
could therefore consider 
\[\label{FlavorHolonomy_g=1}
\rho\left(a_{w}\right)=\rho^{a}x_{a}+\tau\rho^{a}h_{a}+z^{A}\rho^{a}A_{a}+z^{B}\rho^{a}B_{a} \ ,
\]
where $z^{A},z^{B}$ are related to the complex structure of the fibration
and of the base, and $A,B$ are holonomies on the torus. We will not consider this possibility.

The space \eqref{eq:moduli_space_of_flat_connections} has multiple
connected components, some of which correspond to different underlying
bundles on $\Sigma$. The discrete parameters used to identify the
different components can contribute to the discrete data used to define
$D_{oe}$. We have been working under the assumption that all relevant
bundles on the total space $M$ are pullbacks of bundles on the base
$\Sigma$. Moreover, $D_{oe}$ is the pullback of an operator
defined only on $\Sigma$. A bundle on $M$ will be treated as a collection
of complex vector bundles on $\Sigma$. The vector bundles have the
Chern classes specified by \eqref{eq:base_line_bundle_Chern_classes}.
A field charged under $G$ with weight $\rho$ is valued in a line
bundle on $\Sigma$ with first Chern class $\rho^{a}m_{a}$, where
the index $a$ runs over the entire Cartan.

\subsubsection{Gaugino zero modes}

The moduli space of zero modes of the localizing term is in general
a superspace, incorporating the moduli space of flat connections on
$M$ and the fermion zero modes that appear in that background. The
flat $U(N)$-connections we consider, and the $R$-symmetry bundle,
are such that 
\[
{c_{1}}^{2}=c_{2}=0\ ,
\]
and hence do not contribute to the index theorem for the Dirac operator
on $M$. The Dirac operators acting on the gauginos and quarks are also
deformed by the Chern connection, however, this is irrelevant for the
index, as is the imaginary part of the $R$-symmetry connection. The
computation of the Euler number and the signature of $M$ performed
in \cite{Cassani:2013dba} shows that 
\[
\chi\left(M\right)=\sigma\left(M\right)=0\ ,
\]
which is sufficient to determine that the anomaly associated with
the Dirac operator vanishes. We will try to determine the conditions
under which zero modes nevertheless exist for the gauginos.

Consider the supersymmetry equation 
\begin{align}
\delta a_{\mu}=-\frac{1}{2}\left(\epsilon\sigma_{\mu}\tilde{\lambda}+\tilde{\epsilon}\bar{\sigma}_{\mu}\lambda\right)\ .\label{eq:susy_equation_for_connection}
\end{align}
The right hand side defines the fermionic fiber $\hat{\varphi}_{o}$
over the base supermanifold where the equivariant localization takes
place. We will make the simplifying assumption that gaugino zero modes
can only occur in the base manifold $\varphi_{o}$ and hence 
\begin{align}
-\frac{1}{2}\left(\epsilon\sigma_{\mu}\tilde{\lambda}_0+\tilde{\epsilon}\bar{\sigma}_{\mu}\lambda_{0}\right)=0\ .\label{eq:gaugino_zero_mode_restriction}
\end{align}
Then, decomposing $\lambda,\tilde{\lambda}$ as in \eqref{eq:left_handed_decomposition} and contracting \eqref{eq:gaugino_zero_mode_restriction}
with $\bar{K},Y,\bar{Y}$ we get 
\[
\epsilon\lambda_{0}=\tilde{\epsilon}\tilde{\lambda}_{0}=0\ ,
\]
and 
\begin{align}
\frac{\epsilon^{\dagger}\lambda_{0}}{|\epsilon|^{2}}=-\frac{\tilde{\epsilon}^{\dagger}\tilde{\lambda}_{0}}{|\tilde{\epsilon}|^{2}}\equiv a\ .\label{eq:gaugino_zero_mode}
\end{align}
All potential zero modes are of the form 
\[
\lambda_{0}=a\epsilon\ , \qquad\tilde{\lambda}_{0}=-a\tilde{\epsilon}\ .
\]

From the Killing spinor equations, one deduces 
\begin{align}
\begin{aligned}\sigma^{\mu}\left(\nabla_{\mu}+i\left(A_{\mu}+\frac{1}{2}V_{\mu}\right)\right)\tilde{\epsilon} & =0\ ,\\
\bar{\sigma}^{\mu}\left(\nabla_{\mu}-i\left(A_{\mu}+\frac{1}{2}V_{\mu}\right)\right)\epsilon & =0\ .
\end{aligned}
\end{align}
From the equations of motion for the gauginos arising from the $D$-term
for the gauge multiplet we get
\footnote{We neglect the gauge quantum numbers, which we assume have to vanish
in order for the zero mode to arise: the gauginos must be in the same
Cartan as the holonomies.
} 
\begin{align}
\begin{aligned}\sigma^{\mu}D_{\mu}\tilde{\lambda} & =\sigma^{\mu}\left(\nabla_{\mu}+i\left(A_{\mu}-\frac{3}{2}V_{\mu}\right)\right)\tilde{\lambda}=0\ ,\\
\bar{\sigma}^{\mu}D_{\mu}\lambda & =\bar{\sigma}^{\mu}\left(\nabla_{\mu}-i\left(A_{\mu}-\frac{3}{2}V_{\mu}\right)\right)\lambda=0\ .
\end{aligned}
\end{align}
Obviously, for $V=0$, $\lambda_{0}\propto\epsilon$ and $\tilde{\lambda}_{0}\propto\tilde{\epsilon}$
are a solution. 

Assume now that $V\ne0$ and that putative zero modes are defined
by \eqref{eq:gaugino_zero_mode_restriction}. Using the properties of
the supergravity background discussed in Section \ref{ss:Setup}, one can
show that this implies 
\[
\partial_{\mu}a=ia\nabla^{\nu}J_{\nu\mu}\ .
\]
Since a nontrivial solution to a homogenous first order differential
equation on a path connected space is nowhere vanishing, we may write
\[
\nabla^{\mu}\partial_{\mu}\log\, a=i\nabla^{\mu}\nabla^{\nu}J_{\nu\mu}=0\ ,
\]
hence, on a compact manifold, we have 
\[
a=\text{const} \ ,
\]
and 
\[
a\ne0\qquad\Leftrightarrow\qquad\nabla^{\mu}J_{\mu\nu}=0\ .
\]
In turn, $\nabla^{\mu}J_{\mu\nu}=0$ implies that we may choose $V=0$,
and this is possible if and only if $M$ is K{\"a}hler \cite{Closset:2013vra}.
Therefore, with our assumptions, gaugino zero modes exist only for
K{\" a}hler manifolds, in which case they satisfy 
\[
\lambda_{0}\propto\epsilon\ ,\qquad\tilde{\lambda}_{0}\propto\tilde{\epsilon}\ .
\]
We will restrict attention to non-K{\" a}hler manifolds or, equivalently,
$d>0$ in Section \ref{sub:Topology-of-M}.

\section{One-loop determinants and index theorem}\label{ss:IndexTheorem}
In this section, we will evaluate the one-loop partition function \eqref{OneLoopPF}
using the equivariant index theorem for the differential operator $D_{oe}$.
We closely follow the argument of \cite{Pestun:2007rz,Gomis:2011pf,PestunLectureNote}.
The interested reader is referred to \cite{atiyah1974elliptic,atiyah1968indexI,atiyah1968indexII,atiyah1968indexIII} for more details of the Atiyah-Singer index theorem and to \cite{atiyah1967lefschetz,atiyah1968lefschetz,Bott:1987rr} for the Atiyah-Bott localization formula and its applications.

\subsection{Equivariant index theorem}

The one-loop determinants \eqref{OneLoopPF} can be obtained from the $\CR$-equivariant
index of the differential operator 
\begin{align}
\text{ind}(D_{oe})=\tr_{\text{Ker}D_{oe}}e^{\CR}-\tr_{\text{Coker}D_{oe}}e^{\CR}\ .
\end{align}
Once the index is calculated, the partition function is read off from
the weight $w_{\alpha}$ and the multiplicity $c_{\alpha}$ of a representation
$\alpha$ of $\CR$:  
\begin{align}
\text{ind}(D_{oe})=\sum_{\alpha}c_{\alpha}e^{w_{\alpha}}\quad\longrightarrow\quad Z_{\text{one-loop}}=\prod_{\alpha}w_{\alpha}^{-c_{\alpha}}\ .\label{IndRule}
\end{align}
The fields $\varphi_{e}$ and $\varphi_{o}$ are regarded as sections
of bundles $E_{e}$ and $E_{o}$ on a manifold $X$. The differential
operator $D=D_{oe}$ acts on the complex 
\begin{align}
\Gamma(E_{e})\, \xrightarrow{D}\, \Gamma(E_{o})\ .
\end{align}
Let $T=U(1)^{n}$ be the maximal torus of the isometry $\CR$ and
$e^{\CR}=t=(t_{1},t_{2},\cdots,t_{n})$. Using the Atiyah-Singer index
theorem, the index is represented as a sum over the set of fixed points
$F$ of $T$ action:
\begin{align}
\text{ind}_{T}(D)=\sum_{p\in F}\frac{\tr_{E_{e}(p)}t-\tr_{E_{o}(p)}t}{\det_{TX_{p}}(1-t)}\ .\label{AS_index}
\end{align}

To illustrate how it works, consider $X=\BC\BP^{1}$ and the equivariant
index of the Dolbeault operator $D=\bar{\partial}$ acting on the
complex 
\begin{align}
\bar{\partial}:\,\Omega^{0,0}\to\Omega^{0,1}\ ,
\end{align}
under $T=U(1)$ action $z\to tz$ around the fixed point $z=0$ at
the north pole. $\Omega^{0,0}$ and $\Omega^{0,1}$ are generated
by $T$-invariant functions $f(z,\bar{z})$ and $f_{\bar{z}}(z,\bar{z})d\bar{z}$, around the north pole. Since under $U(1)$ action $f\to f$
and $f_{\bar{z}}\to tf_{\bar{z}}$, we obtain $\tr_{\Omega^{0,0}}t=1$
and $\tr_{\Omega^{0,1}}t=t$. The tangent bundle $TX$ is generated
by $\partial_{z}$ and $\partial_{\bar{z}}$ with $T$ eigenvalues
$t^{-1}$ and $t$, and $\det_{TX_{z=0}}(1-t)=(1-t)(1-t^{-1})$. Then
the north pole $z=0$ contributes to the index \eqref{AS_index} by
\begin{align}
\text{ind}_{T}(\bar{\partial})|_{z=0}=\frac{1}{1-t^{-1}}=\sum_{k=0}^{\infty}t^{-k}\ .\label{Indz=00003D0}
\end{align}
This result can be understood as a counting of $U(1)$ invariant holomorphic
functions on $\BC$ which is the kernel of the Dolbeault operator
$\bar{\partial}$ 
\begin{align}
f(z)=\sum_{k=0}^{\infty}c_{k}z^{k}\ .
\end{align}
Under $U(1)$ action $z\to tz$, the coefficients transform as $c_{k}\to t^{-k}c_{k}$
so as to $f(z)=f(tz)$. The index is nothing but the summation of
the weight $t^{-k}$ of the holomorphic functions.

On the other hand, the other fixed point at the south pole $z=\infty$
can be treated by introducing a different patch $w=1/z$. In this
patch, $\Omega^{0,0}$ and $\Omega^{0,1}$ are generated by $f(w,\bar{w})$
and $f_{\bar{w}}(w,\bar{w})d\bar{w}$
, and the tangent bundle $TX$ is generated by $\partial_{w}$ and
$\partial_{\bar{w}}$ with $T$ eigenvalues $t$ and $t^{-1}$. Thus
the contribution to the index from the south pole $w=0$ is 
\begin{align}
\text{ind}_{T}(\bar{\partial})|_{w=0}=\frac{1}{1-t}\ .
\end{align}
The sum of the two gives the total index 
\begin{align}
\text{ind}_{T}(\bar{\partial})=\frac{1}{1-t^{-1}}+\frac{1}{1-t}=1\ .\label{IndS2}
\end{align}

We can twist the Dolbeault complex by the holomorphic line bundle
$\CO(n)$ with the first Chern class $c_{1}=n$. Now the complex is
\begin{align}
\bar{\partial}:\,\Omega^{0,0}(\CO(n))\to\Omega^{0,1}(\CO(n))\ .
\end{align}
We define the action of $T$ on the fiber of $\CO(n)$ in the $z$
patch to be $t^{n/2}$. In this patch, $\Omega^{0,0}(\CO(n))$
and $\Omega^{0,1}(\CO(n))$ are generated by $\phi(z)$ and $\phi_{\bar{z}}(z)d\bar{z}$
with $T$ eigenvalues $t^{n/2}$ and $t^{1+n/2}$ for $\phi(z)$ and $\phi_{\bar{z}}(z)$. The index at $z=0$ is $t^{n/2}$ times the untwisted index \eqref{Indz=00003D0}
\begin{align}
\text{ind}_{T}(\bar{\partial};\CO(n))|_{z=0}=\frac{t^{n/2}}{1-t^{-1}}\ .
\end{align}
In the $w$-patch, a section $\phi\in\Omega^{0,0}(\CO(n))$ transforms
under the coordinate change by 
\begin{align}
\phi(z)=z^{n}\tilde{\phi}(w)\ .
\end{align}
It follows that $\Omega^{0,0}(\CO(n))$ and $\Omega^{0,1}(\CO(n))$
are generated by $\tilde{\phi}(w)$ and $\tilde{\phi}_{\bar{w}}(w)d\bar{w}$
with $T$ eigenvalues $t^{-n/2}$ and $t^{-1-n/2}$ for $\tilde{\phi}(w)$
and $\tilde{\phi}_{\bar{w}}(w)$. The index from the south pole is
\begin{align}
\text{ind}_{T}(\bar{\partial};\CO(n))|_{w=0}=\frac{t^{-n/2}}{1-t}\ ,
\end{align}
and the total index is 
\begin{align}
\begin{aligned}\text{ind}_{T}(\bar{\partial};\CO(n)) & =\frac{t^{n/2}}{1-t^{-1}}+\frac{t^{-n/2}}{1-t}\ ,\\
 & =\left\{ \begin{array}{cl}
t^{n/2}\sum_{k=0}^{n}t^{-k}\ , & n\ge0\ ,\\
0\ , & n=-1\ ,\\
-t^{n/2}\sum_{k=0}^{-n-2}t^{k+1}\ , & n<-1\ .
\end{array}\right.
\end{aligned}
\label{IndS2LineBundle}
\end{align}

A Dirac operator $D_{\text{Dirac}}$ acting on spinor bundles 
\begin{align}
D_{\text{Dirac}}:S^{+}\to S^{-}\ ,
\end{align}
is isomorphic to the Dolbeault complex by twisting by the square root
of the canonical bundle $\CK$ on K{\" a}hler manifolds 
\begin{align}
D_{\text{Dirac}}=\frac{1}{2}(\bar{\partial}+\bar{\partial}^{\ast}):\Omega^{0,\text{even}}(X,\CK^{1/2})\to\Omega^{0,\text{even}}(X,\CK^{1/2})\ .
\end{align}
The Dirac operator on $\BC\BP^{1}$ of the twisted complex 
\begin{align}
D_{\text{Dirac}}:S^{+}\otimes\CO(n)\to S^{-}\otimes\CO(n)\ ,
\end{align}
is equal to the Dolbeault complex 
\begin{align}
\bar{\partial}:\,\Omega^{0,0}(\CO(n)\otimes \CK^{1/2})\to\Omega^{0,1}(\CO(n)\otimes \CK^{1/2})\ .
\end{align}
and the equivariant index is similarly calculated as 
\begin{align}
\begin{aligned}\text{ind}_{T}(D_{\text{Dirac}};\CO(n)) & =t^{-1/2}\frac{t^{n/2}}{1-t^{-1}}+t^{1/2}\frac{t^{-n/2}}{1-t}\\
 & =\frac{t^{-1/2}(t^{n/2}-t^{-n/2})}{1-t^{-1}}\ ,
\end{aligned}
\label{DiracIndex}
\end{align}
where the factor $t^{-1/2}$ ($t^{1/2}$) comes from the canonical
bundle at $z=0$ ($w=0$).

Next we consider a manifold $X$ on which a compact Lie group $U$
acts freely. Let $Y=X/U$ be the quotient and $\pi:X\to Y$ be the
associated $U$-principal bundle. Given a $T$-equivariant operator
$D_{Y}$ for a complex of vector bundles $E_{Y}$ on $Y$, a $U\times T$
equivariant operator $D_{X}$ and a complex of vector bundles $E_{X}$
are obtained as pullbacks by $\pi^{\ast}$: 
\begin{align}
E_{X}=\pi^{\ast}E_{Y}\ ,\qquad D_{X}=\pi^{\ast}D_{Y}\ .
\end{align}
We can compute the $U\times T$ equivariant index for the complex
$(E_{X},D_{X})$ by using the index on $Y$ as 
\begin{align}
\text{ind}_{U\times T}(D_{X})=\sum_{\alpha\in R_{U}}\text{ind}_{T}(D_{Y}\otimes W_{\alpha})\, \chi_{\alpha}\ ,\label{ASIndFree}
\end{align}
where $R_{U}$ is the set of irreducible representations of $U$,
$\chi_{\alpha}$ the character of the representation $\alpha$, and
$W_{\alpha}$ the vector bundle over $Y$ associated to the $U$-principal
bundle.

Let us apply the index formula to our four-manifold $M$ with $U=U(1)^{2}$
and the base Riemann surface $Y=\Sigma$. Irreducible representations
of $U$ are parametrized by two integers $\alpha=(n,l)$ and the character
is $\chi_{\alpha}=x^{n}y^{l}$ where $x,y$ are constant. The vector
bundle $W_{\alpha}$ depends on how $U=U(1)^{2}$ is fibered over
$Y$. We consider the Dirac operator as a $T$-equivariant operator
$D_{Y}=D_{\text{Dirac}}$ for the matter sector. Then the index formula
\eqref{ASIndFree} yields the index for a $U\times T$ equivariant
operator $D_{X}$ on $X$ 
\begin{align}
\text{ind}_{U\times T}(D_{X})=\sum_{n,l\in\BZ}\text{ind}_{T}(D_{\text{Dirac}}\otimes W_{n,l})\, x^{n}y^{l}\ .\label{T2S2}
\end{align}

\subsection{Lens space}

A torus fibration over a Riemann surface $\Sigma$ is characterized
by two first Chern classes for each circle. We consider
the case where one of the circle is nontrivially fibered over a two-sphere,
i.\,e., $M=S^{1}\times L(d,1)$ with the lens space $L(d,1)$.
Since the fibration is nontrivial, the vector bundle $W_{\alpha}$
is the line bundle $\CO(d l)$ over $Y=S^2$, where we choose
$y$ to be the equivariant parameter for the $U(1)$ fiber of degree
$d$ for the lens space.

\subsubsection{Matter sector}
The Dirac operator of the matter sector acts on the fermion of $R$-charge
$r-1$ and the gauge representation $\rho$.
The integers $n$ and $l$ are physically interpreted as the Kaluza-Klein momenta along the trivial circle and the nontrivial fiber circle.
 The twisted complex is given by
\begin{align}\label{MatterDiracOp}
D_{\text{Dirac}}:S^{+}\otimes\CO(dl + \rho(m))\otimes  L^{r-1}\otimes E_{\rho}~\to~S^{-}\otimes\CO(dl + \rho(m))\otimes L^{r-1}\otimes E_{\rho}\ ,
\end{align}
where $L$ and $E_{\rho}$ are the $R$-symmetry  line bundle and the gauge bundle of $\rho$ representation, respectively. 
We also take into account the effect of holonomy which shift the degree of the circle line bundle by $\rho(m)$. 
We already know the equivariant index of the Dirac operator and we can calculate
and rewrite it as follows: 
\begin{align}
\begin{aligned}\text{ind}_{G\times U\times T}(D_{\text{matter}}) & =\sum_{n,l\in\BZ}t^{-r/2}\frac{t^{(dl+\rho(m))/2}-t^{-(dl+\rho(m))/2}}{1-t^{-1}}x^{n}y^{dl+\rho(m)}u\ ,\\
 & =\sum_{n,l\in\BZ}x^{n}u\,(pq)^{r/2}\,\frac{q^{-(dl+\rho(m))}-p^{dl+\rho(m)}}{1-pq}\ ,
\end{aligned}
\end{align}
where we introduced the new variables $p,q$ 
\begin{align}
t=(pq)^{-1}\ ,\qquad y=(p/q)^{1/2}\ ,\label{Var_pq}
\end{align}
and $u=\rho(g)$ with the equivariant parameter $g$ for the gauge
symmetry. In the first line, we multiplied $t^{-(r-1)/2}$ as the
index of the $R$-symmetry bundle for a fermion of $R$-charge
$r-1$.

To encode the index to the one-loop partition function with \eqref{IndRule},
we employ a dictionary between the elliptic gamma function and the
equivariant index 
\begin{align}
\sum_{n,l\in\BZ}x^{n}u\frac{p^{l}}{1-q}\quad\leftrightarrow\quad e^{i\pi\CE(u/p,1/p,q)}\Gamma(u;p,q)^{-1}\ ,
\end{align}
where $\CE$ is a phase factor arising from the regularization of the infinite product \cite{ruijsenaars2000barnes,felder2000elliptic,friedman2004shintani,Closset:2013sxa,Assel:2014paa} 
\begin{align}
\CE(u,p,q)=\frac{w^{3}}{3\tau\sigma}+\frac{2-\tau^{2}-\sigma^{2}}{12\tau\sigma}w\ ,\qquad w=z -\frac{\tau+\sigma}{2}\ ,
\end{align}
with $p=e^{2\pi i\sigma},q=e^{2\pi i\tau}$ and $u=e^{2\pi iz}$.
One may confirm this by rewriting the infinite summation and expanding
$1/(1-q)$ as follows 
\begin{align}
\begin{aligned}\sum_{n\in\BZ}x^{n}u \sum_{i\in\BZ}\frac{p^{i}}{1-q} & =\sum_{n\in\BZ}x^{n}u\sum_{i=0}^{\infty}\left[p^{i}\frac{q^{-1}}{q^{-1}-1}+p^{-(i+1)}\frac{1}{1-q}\right]\ ,\\
 & =\sum_{n\in\BZ}x^{n}u\sum_{i,j=0}^{\infty}\left[p^{-(i+1)}q^{j}-p^{i}q^{-(j+1)}\right]\ ,\\
 & \leftrightarrow\quad e^{i\pi\CE(u/p,1/p,q)}\,\Gamma(z/p;1/p,q)\ .
\end{aligned}
\end{align}
In the final line, we used the rule \eqref{IndRule} between the index and the one-loop determinant, that is  
\begin{align}
\sum_{n\in\BZ}\sum_{i,j=0}^{\infty}x^{n}u\left[p^{i}q^{j}-p^{-(i+1)}q^{-(j+1)}\right]\qquad\leftrightarrow\qquad e^{i\pi\CE(u,p,q)}\Gamma(u;p,q)\ ,
\end{align}
where $\Gamma$ is the elliptic Gamma function defined by \eqref{DefEllipticGamma}.

It follows that the one-loop partition function of a chiral multiplet of $R$-charge $r$ 
on $S^{1}\times L(d,1)$ is 
given by
\begin{align}\label{LensIndexMatter}
Z_{\text{matter}}^{(r,\rho)}(m,u)& =
e^{i\pi\CE^{(r)}(\rho(m),u)}\,\Gamma(u(pq)^{r/2}q^{d-\rho(m)};q^{d},pq)\,\Gamma(u(pq)^{r/2}p^{\rho(m)};p^{d},pq)\ ,
\end{align}
with 
\begin{align}
\CE^{(r)} (m,u)=\CE(u(pq)^{r/2}q^{d - m},q^{d},pq)-\CE(u(pq)^{r/2}p^{m-d},p^{-d},pq)\ .
\end{align}
This agrees with the lens index obtained in \cite{Benini:2011nc,Razamat:2013opa}
by the orbifold projection up to the phase factor.
To make contact with their results, the phase factor can be cast into
\begin{align}\label{ZeroPointEnergy}
e^{i\pi \CE^{(r)}(\rho(m),u)} = e^{i\pi \CE^{(r)}_0(u)}\, \CI_0^{(r)} (\rho(m), u) \ ,
\end{align}
where $\CI_0^{(r)}$ is the ``zero point energy'' depending on the holonomies $m$ that appeared  in \cite{Benini:2011nc,Razamat:2013opa}
\begin{align}
\begin{aligned}
	\CI_0^{(r)} (m, u) = \left( (pq)^\frac{1-r}{2} u^{-1} \right)^\frac{m(d - m)}{2d} \left( \frac{p}{q} \right)^\frac{m(d - m)(d - 2m)}{12d} \ ,
\end{aligned}
\end{align}
and $\CE_0^{(r)}(u)$ is the remaining phase independent of $m$
\begin{align}
\CE_0^{(r)}(u) = \frac{(2 z +(r-1) (\sigma +\tau )) \left(4 z^2 + 2d^2 \sigma  \tau +r^2
   (\sigma +\tau )^2 + 4 z(r-1) (\sigma +\tau ) - 2 r(\sigma +\tau )^2+2\right)}{24d \sigma  \tau } \ ,
\end{align}
if $u=e^{2\pi i z}$.
When $m=0$, the zero point energy $\CI_0^{(r)}$ vanishes and only the ``supersymmetric Casimir energy" $\CE_0^{(r)}$ remains \cite{Assel:2014paa}.

\subsubsection{Gauge sector}

The relevant differential operator for the gauge sector is the de
Rham operator whose complex is 
\begin{align}
d:\Omega^{0}\,\xrightarrow{d}\,\Omega^{1}\,\xrightarrow{d}\,\Omega^{2}\ .
\end{align}
Here we consider the equivariant index on $X=S^{2}$ with respect
to $T=U(1)$ acting on the complex coordinate $z$ around the north
pole as $z\to tz$. The complexification of the de Rham complex is
isomorphic to the Dirac complex 
\begin{align}
D_{\text{Dirac}}=d+d^{\ast}:\Omega^{1}\to\Omega^{0}\oplus\Omega^{2}\ ,
\end{align}
namely, the relation between the indices is $\text{ind}_{T}(d)=-\text{ind}_{T}(D_{\text{Dirac}})$.

The Dirac operator acts on the gaugino of $R$-charge $-1$ in the vector multiplet.
The complex is obtained from that of the matter sector in \eqref{MatterDiracOp} by setting $r=0$ and replacing the representation $\rho$ with an adjoint representation $\alpha$.
Therefore, the index of the gauge sector is equal to the minus of the index of the matter sector with the replacements:
\begin{align}\label{ReplaceGM}
\text{ind}_{G\times U\times T}(D_{\text{gauge}}) & =  - \text{ind}_{G\times U\times T}(D_{\text{matter}}) \big|_{r=0, \rho\to \alpha} \ .
\end{align}

Combining with the relation \eqref{ReplaceGM}, we end up with the one-loop partition function of the gauge sector
\begin{align}\label{GaugeLensPF}
\begin{aligned}
Z_\text{gauge}(m) &=  \prod_{\alpha\in \text{Ad}\,G} e^{-i\pi \CE^{(0)}(\alpha(m), v)} \left(\Gamma(v q^{d-\alpha(m)};q^{d},pq)\Gamma(vp^{\alpha(m)} ;p^{d},pq)\right)^{-1}\ , 
\end{aligned}
\end{align}
where $v\equiv\alpha(g)$ and $p,q$ are defined in \eqref{Var_pq}.
$\text{Ad}\,G$ is the adjoint representation of the gauge group.
Again, this is equal to the lens index of the gauge sector \cite{Benini:2011nc,Razamat:2013opa} up to the phase factor because of the relation \eqref{ZeroPointEnergy} between the phase $\CE^{(0)}$ and the zero point energy.\footnote{In addition to the phase $\CE^{(0)}$, our partition function \eqref{GaugeLensPF} differs from that of the literatures \cite{Benini:2011nc,Razamat:2013opa} by a term which arises from the gauge fixing. We compensate the term by the measure of the matrix model.}

\subsection{$T^{2}\times S^{2}$}
Let us make comments on the case with $g=0$ and $d=0$, i.e., $M=T^2\times S^2$.
This manifold is K{\"a}hler and there exists gaugino zero modes that prevents us from the complete analysis of the partition function as we will describe below.

On $T^{2}\times S^{2}$, the metric is given by \eqref{T2OverRS}
with 
\begin{align}
\Omega=1\ ,\qquad h=\bar{h}=0\ ,\qquad c=\frac{2}{1+|z|^{2}}\ .
\end{align}
The $R$-symmetry background gauge field \eqref{RChern} has nontrivial
field strength through $S^{2}$ 
\begin{align}
A^{c}=\frac{i}{2}\frac{\bar{z}dz-zd\bar{z}}{1+|z|^{2}}+\frac{i}{2}d\log s\ ,
\end{align}
whose first Chern class is $c_{1}(L)=-1$ as is consistent with \eqref{eq:R_symmetry_Chern_class} in the discussion of Section \ref{ss:CSandRsym}. 
Namely, the $R$-symmetry line bundle is a line bundle $\CO(-1)$ of degree $-1$.

For the matter sector, we consider the Dirac operator on $S^{2}$ acting on a fermionic field of $R$-charge $r-1$ 
\begin{align}
D_{\text{Dirac}}:S^{+}\otimes\CO(-(r-1))\otimes E_\rho \to S^{-}\otimes\CO(-(r-1))\otimes E_\rho \ .
\end{align}
The total index of the matter sector on $T^{2}\times S^{2}$ reads 
\begin{align}
\begin{aligned}\text{ind}_{G\times U\times T}(D_{X}) & =\sum_{n,l\in\BZ}\frac{t^{-r/2}-t^{-1+r/2}}{1-t^{-1}}x^{n}y^{l}\rho(g)\ ,\\
 & =\left\{ \begin{array}{cl}
\sum_{n,l\in\BZ}\sum_{k=-\frac{|r|}{2}}^{\frac{|r|}{2}}t^{k}x^{n}y^{l}\rho(g)\ , & r\le0\ ,\\
0\ , & r=1\ ,\\
-\sum_{n,l\in\BZ}\sum_{k=-\frac{r}{2}+1}^{\frac{r}{2}-1}t^{k}x^{n}y^{l}\rho(g)\ , & r>1\ ,
\end{array}\right.
\end{aligned}
\end{align}
where the $R$-charge $r$ is quantized to be integer \cite{Closset:2013sxa}.
Decoding it with the rule \eqref{IndRule}, we obtain the one-loop
partition function of the matter sector on $T^{2}\times S^{2}$ 
\begin{align}\label{MatterPFT2S2}
\begin{aligned}
Z_{\text{matter}}^{(r,\rho)} & =\left\{ \begin{array}{cl}
\prod_{k=-\frac{|r|}{2}}^{\frac{|r|}{2}}\prod_{n,l\in\BZ}(2\pi i \xi)^{-1}\left( n+l \tau  +k \sigma + \rho(a)\right)^{-1}\ , & r\le0\ ,\\
\prod_{k=-\frac{r}{2}+1}^{\frac{r}{2}-1}\prod_{n,l\in\BZ} (2\pi i \xi)\left(n+ l\tau +k \sigma +\rho(a)\right)\ , & r>1\ ,
\end{array}\right.\\
 & =\left\{ \begin{array}{cl}
\prod_{k=-\frac{|r|}{2}}^{\frac{|r|}{2}}-i \frac{\eta(\tau)}{\vartheta_{1}\left(k\sigma+\rho\left(a\right)|\tau\right)}\ , & r\le0\ ,\\
\prod_{k=-\frac{r}{2}+1}^{\frac{r}{2}-1} i\frac{\vartheta_{1}\left(k\sigma+\rho\left(a\right)|\tau\right)}{\eta(\tau)}\ , & r>1\ ,
\end{array}\right.
\end{aligned}
\end{align}
where the new parameters are introduced by\footnote{In the second equality of \eqref{MatterPFT2S2}, we throw away the factor $\prod_{n,l\in\BZ}(2\pi i\xi)^{\pm 1}$ which would
become one after the zeta-function regularization.}
\begin{align}\label{T2S2Parameters}
\begin{aligned}t & =e^{2\pi i\hat{\sigma}}\ ,&\qquad x&=e^{2\pi i\xi}\ ,&\qquad y&=e^{2\pi i\hat{\tau}}\ ,\qquad g=e^{2\pi i\hat{a}}\ ,\\
\tau & =\frac{\hat{\tau}}{\xi}\ ,&\qquad\sigma&=\frac{\hat{\sigma}}{\xi}\ ,&\qquad a&=\frac{\hat{a}}{\xi}\ ,
\end{aligned}
\end{align}
and the infinite products are regularized with the identity \cite{quine1993zeta}
\begin{align}
\begin{aligned}\prod_{n,l\in\BZ}(n+ l \tau +z) & =i \frac{\vartheta_{1}(z|\tau)}{\eta(\tau)}\ .
\end{aligned}
\end{align}

Our results \eqref{MatterPFT2S2} for the matter sector agree with those of \cite{Closset:2013sxa} 
that are derived by reducing the four-dimensional theory to $\CN = (0,2)$ theories on $T^2$.
They point out that the partition functions for $r> 1$ and $r\le 0$ are the contribution from the Fermi and chiral multiplets of the $\CN = (0,2)$ theories, respectively.

The one-loop partition function of the gauge sector on $T^{2}\times S^{2}$ is obtained similarly by using the result \eqref{MatterPFT2S2} for the matter sector and the relation \eqref{GaugeLensPF}
\begin{align}\label{GaugePFT2S2}
Z_{\text{gauge}}  & =\prod_{\alpha\in\text{Ad}\,G}i \frac{\vartheta_{1}\left(\alpha(a)|\tau\right)}{\eta(\tau)}\ ,
\end{align}
where $\eta$ and $\vartheta_1$ are the Dedekind's eta and Jacobi's theta functions defined by \eqref{etaDef} and \eqref{JacobiDef}.
Naively, the gauge sector partition function \eqref{GaugePFT2S2} can be interpreted as
contributions of the vector multiplets
of $\CN=(0,2)$ theories in two-dimensions \cite{Benini:2013nda,Benini:2013xpa}.

The partition function \eqref{GaugePFT2S2}, however, is zero because $\vartheta_1(\alpha(a)|\tau)$ vanishes for the Cartan generators with $\alpha = 0$.
Also, there are the gaugino zero modes on $T^2\times S^2$, which our derivation assumed not to exist so far.
To fix this, we go back to the infinite product form \eqref{MatterPFT2S2} for the Cartan generators, remove them and use the zeta-function regularization
\begin{align}
\prod_{n,l\in \BZ, \,n,l\neq 0} (n + l\tau) = 2\pi i\, \eta^2 (\tau) \ .
\end{align}
Thus the one-loop partition function of the gauge sector is given by
\begin{align}\label{GaugePFT2S2Full}
Z_{\text{gauge}}  & = \left( 2\pi i \eta^2 (\tau)\right)^{\text{rank}\, G}\prod_{\alpha\in\text{Ad}\,G} i\frac{\vartheta_{1}\left(\alpha(a)|\tau\right)}{\eta(\tau)}\ .
\end{align}
The gaugino zero modes $\lambda_0$ and $\tilde\lambda_0$ from $n=l=\alpha=0$ give rise to the measure $\CD \lambda_0 \CD\tilde\lambda_0$ that should be taken into account in the path integral.

The one-loop partition function for the matter sector \eqref{MatterPFT2S2} is no longer legitimate in the presence of the gaugino zero modes because of the interaction of the form $\tilde\phi \lambda_0\psi$. 
To simplify the discussion, let us consider a rank-one gauge theory and try to write down the resulting partition function on $T^2\times S^2$.
After localizing the gauge sector, we end up with the one-loop partition function $Z_\text{gauge}$ and the measure of the path integral given by the complex holonomy $a$ around the torus.
Then, using the localizing action \eqref{Loc_Mat} for the matter of $R$-charge $r$ and gauge charge $q$, the resulting partition function takes the form
\begin{align}
\begin{aligned}
Z_{T^2\times S^2} &\sim  \int d^2 a  \, Z_\text{gauge} (a)
\int \CD \lambda_0 \CD \tilde\lambda_0\, \CD\phi \CD\psi\CD\tilde\psi\, e^{-S_\text{loc}(\phi, \psi,\tilde\psi, \lambda_0, \tilde\lambda_0)}\ ,
\end{aligned}
\end{align}
with the matter action around the fixed points \eqref{Mat_Locus}\footnote{We use the fact that $V_\mu=0$ and $R=-2$ for $T^2\times S^2$}
\begin{align}
S_\text{loc} = \int_{T^2\times S^2} d^4 x \sqrt{g} \left[ D_{\mu}\phi^\dagger D^{\mu}\phi + \frac{r}{2}\tilde{\phi} \phi
+\frac{1}{2}\psi^\dagger\bar{\sigma}^{\mu}D_{\mu}\psi - i\phi^\dagger \lambda_0 \psi + i \tilde\psi \tilde\lambda_0 \phi\right]\ .
\end{align}
Decomposing the fields with the spherical harmonics of $S^2$, $S_\text{loc}$ gives actions of long, chiral and Fermi multiplets of $\CN=(0,2)$ theories on $T^2$ independent of each other.
Since the gaugino zero modes make the path integral subtle, we have to carry out the path integral along the line of the careful analysis in \cite{Benini:2013nda,Benini:2013xpa} where the auxiliary $D$ field in the gauge sector is kept as a regulator by the end of computations. 
We will not pursue this interesting issue in this paper and leave it for future investigation.

\subsection{Elliptic fibration over Riemann surface}

We have considered the case with base $S^{2}$ so far. Now we move
to Riemann surfaces with genus $g\ge1$ where there is no $U(1)$
equivariant symmetry for $g\ge2$. Thus, we use the usual Atiyah-Singer
index theorem instead of the equivariant one.

The index of the Dirac operator $D_{\text{Dirac}}:S^{+}\otimes E\to S^{-}\otimes E$
acting on the spinor bundles twisted by a vector bundle $E$ is given
by 
\begin{align}
\text{ind}(D_{\text{Dirac}};E)=\int_{X}\hat{A}(TX)\,\text{ch}(E)\ ,
\end{align}
where $\hat{A}$ and $\text{ch}$ are the $A$-roof genus and the
Chern character. The formula reduces on Riemann surfaces
$X=\Sigma$ to 
\begin{align}
\text{ind}(D_{\text{Dirac}};E)=\int_{\Sigma}1\cdot c_{1}(E)=\text{deg}(E)\ .
\end{align}

For the matter sector, the complex is twisted by the $R$-symmetry
line bundle $L$, which has the first Chern class \eqref{eq:R_symmetry_Chern_class}
\begin{align}
c_{1}(L)=-\frac{\chi(\Sigma)}{2}=g-1~~\text{mod}\, d\ .
\end{align}
The Dirac operator acts on the fermion of $R$-charge $r-1$ in $\rho$ representation \eqref{MatterDiracOp}, whose index is inferred as 
\begin{align}
\begin{aligned}
\text{ind}(D_{\text{matter}})=\sum_{\rho\in\FR}\sum_{n,l\in\BZ}\left(-(r-1)\frac{\chi(\Sigma)}{2}+dl + \rho(m)\right)x^{n}y^{dl - (r-1)\frac{\chi(\Sigma)}{2}}u\ ,
\end{aligned}
\end{align}
where the shift of the exponent of $y$ comes from the holonomy along the fiber direction of the $R$-symmetry line bundle \eqref{R_bundle_holonomy}.
The one-loop partition function follows as
\begin{align}
\begin{aligned}
Z_{\text{matter}}^{(r,\rho)} &=\prod_{\rho\in\FR}\prod_{n,l\in\BZ}\left(n+\tau d\left( l - (r-1)\frac{\chi(\Sigma)}{2d}\right) +\rho(a_w) \right)^{-(r-1)\frac{\chi(\Sigma)}{2}+dl + \rho(m)}\ , 
\end{aligned}
\end{align}
where we used the same parametrization \eqref{T2S2Parameters}.
$\rho(a_w)$ is the holonomy given by \eqref{FlavorHolonomy_g>1} for $g>1$ and \eqref{FlavorHolonomy_g=1} for $g=1$, respectively.

Similarly for the gauge sector, the Dirac operator acts on the gaugino
of $R$-charge $-1$, and the one-loop partition function is given by
\begin{align}
Z_{\text{gauge}}=\prod_{\alpha\in \text{Ad}\, G} \prod_{n,l\in\BZ}\left(n+\tau d\left( l +\frac{\chi(\Sigma)}{2d}\right) +\rho(a_w) \right)^{\frac{\chi(\Sigma)}{2}+dl + \rho(m)} \ .
\end{align}

\section{The generalized index}\label{ss:Generalized_Index}

The result for the partition function on $M$ is a function of the
following parameters:
\begin{enumerate}
\item $g$ - the genus of $\Sigma$, and $d$ - the Chern class of the circle
bundle $M_{3}$. As usual $\chi\left(\Sigma\right)=2-2g$.
\item The part of the complex structure moduli space of $M$ specified by
the complex numbers $\sigma$ and $\tau$. In the case $g=0$ and
$d\ge3$ there is also a discrete choice between complex structures
I and II. 
\item The gauge group $G$ and the matter representation $\mathfrak{R}$
in the group $G\times F$, where $F$ is the flavor symmetry group.
\item A choice of non-anomalous $R$-symmetry under which a chiral superfield
has a charge denoted by $r$. The restriction on $r$ is 
\[
r\left(-\frac{\chi\left(\Sigma\right)}{2}\text{ mod }d\right)\in\mathbb{Z}\ .
\]
This does not apply to $g=0$ in the special component I. We denote by $\chi_{d}\left(\Sigma\right)$ the unique integer representing 
\[
-\frac{\chi\left(\Sigma\right)}{2}\text{ mod }d\ ,
\] 
in the range $0,\ldots,d-1$.
\item For every $U(1)$ factor of $G$, a Fayet-Iliopoulos term $\xi$ which
may be quantized. This also requires a choice of element $W\in H^{1,2}\left(M\right)$
which determines $\kappa$. We denote 
\[
\kappa_{M}=\int_{M}\sqrt{g}\,\kappa\ .
\]
We describe this contribution in the next subsection.
\item The moduli space of flat $F$-connections on $M$.
\end{enumerate}

We state the final result only for $G=U(N)$. 

\begin{itemize}
\item $g=0$: Defining $p=e^{2\pi i\sigma},q=e^{2\pi i\tau}$ and $u=e^{2\pi iz}$,
\begin{align}\label{LensIndexMatter-1}
\begin{aligned}
Z_{\text{matter}}^{(r,\rho)}  \left(z,m\right)=&\, e^{i\pi\left( \CE(u(pq)^{r/2}q^{d-m},q^{d},pq)-\CE(u(pq)^{r/2}p^{m-d},p^{-d},pq)\right)} \\
	&\quad \cdot\Gamma(u(pq)^{r/2}q^{d-m};q^{d},pq)\,\Gamma(u(pq)^{r/2}p^{m};p^{d},pq)\ , \\
Z_{\text{gauge}}  \left(z,m\right)=&\, e^{-i\pi\CE^{(0)}(\alpha(m),v)}\left(\Gamma(uq^{d-m};q^{d},pq)\Gamma(up^{m};p^{d},pq)\right)^{-1}\ ,\end{aligned}
\end{align}
with 
\begin{align}
\CE(u,p,q)=\frac{w^{3}}{3\tau\sigma}+\frac{2-\tau^{2}-\sigma^{2}}{12\tau\sigma}w\ ,\qquad w=z-\frac{\tau+\sigma}{2}\ ,
\end{align}
and
\begin{align}
\CE^{(0)} (m,u)=\CE(uq^{d - m},q^{d},pq)-\CE(up^{m-d},p^{-d},pq)\ .
\end{align}
Here, $z$ and $m$ will specify a connection, consisting of a continuous holonomy around the $S^1$ and a discrete holonomy on the fiber respectively, on the group $G\times F$. The summation and integration are over the gauge part of this connection, $x$ and $\tilde{m}$, and the flavor part is denoted as $a_{f}$. The partition function is given by
\begin{align}\label{eq:final_partition_function_lens}
\begin{aligned}
Z_{0,d}\left(\sigma,\tau,\xi,a_{f}\right)&=\\
\frac{1}{|\CW|}\int_{\mathcal{M}_{G}^{0}\left(0,d\right)}&e^{-i\xi a^{U(1)}_{w}\kappa_{M}}
\prod_{\alpha\in\text{Ad}\,G}Z_{\text{gauge}}\left(\alpha\left(z\right),\alpha\left(m\right)\right)\prod_{\rho\in\mathfrak{R}_{G\times F}}Z_{\text{\text{matter}}}^{\left(r,\rho\right)}\left(\rho\left(z\right),\rho\left(m\right)\right) \ ,
\end{aligned}
\end{align}
where
\[
\int_{\mathcal{M}_{G}^{0}\left(0,d\right)}=\prod_{a=1}^{N}\left(\sum_{\tilde{m}_{a}\in0,\ldots, d-1}\int_{0}^{1}\frac{x_{a}}{2\pi}\right) \ .
\]
and 
\[
|\CW| = \prod_{l} n_{l}! \ ,
\]
 is the residual Weyl factor. In this case $n_l$ are the multiplicity of the different values for $m_a$.

The partition function above is applicable in the complex structure I, the one usually used to define the lens space. To recover the partition function in the complex structure II we need only to replace $m$ by $m+(r-1)(d-1)$ in the matter sector, and by $m+1$ in the gauge sector. This replacement takes into account the additional $R$-symmetry flux on $\Sigma$.
\end{itemize}

\begin{itemize}
\item $g\ge1$:
Here, $a_{w}$ will denote a connection on the group $G\times F$. The summation and integration are over the gauge part, and the flavor part is denoted as $a_{f}$. An index $a$ into the Cartan is such that $x_{a}=x_{j\left(a\right)}$.
For a given weight $\rho\in\mathfrak{R}$, we define
\[
\rho\left(a_{w}\right)=\rho^{a}\left(x_{a}+\tau\frac{m_{a}}{dN_{j\left(a\right)}}\right)\ ,\qquad\rho\left(m\right)=\rho^{a}m_{a}\ ,
\]
\begin{align}
\begin{aligned}
Z_{\text{matter}}^{\left(r, \rho\right)} & \left(z,m\right)=\prod_{n,l\in\BZ}\left(n+\tau\left(l+(r-1)\frac{\chi_{d}(\Sigma)}{d}\right)+z\right)^{\left(r-1\right)\chi_{d}(\Sigma)+dl+m}\ ,\\
Z_{\text{gauge}} & \left(z,m\right)=\prod_{n,l\in\BZ}\left(n+\tau\left(l-\frac{\chi_{d}(\Sigma)}{d}\right)+z\right)^{-\chi_{d}(\Sigma)+dl+m}\ .
\end{aligned}
\end{align}
The partition function is given by
\begin{align}\label{eq:final_partition_function}
\begin{aligned}
Z_{g,d}\left(\tau,\xi,a_{f}\right)=&\\
\frac{1}{|\CW|}\int_{\mathcal{M}_{G}^{0}\left(g,d\right)}
&e^{-i\xi a^{U(1)}_{w}\kappa_{M}}\prod_{\alpha\in\text{Ad}\,G}Z_{\text{gauge}}\left(\alpha\left(a_{w}\right),\alpha\left(m\right)\right)\prod_{\rho\in\mathfrak{R}_{G\times F}}Z_{\text{\text{matter}}}^{\left(r,\rho\right)}\left(\rho\left(a_{w}\right),\rho\left(m\right)\right) \ ,
\end{aligned}
\end{align}
where, in the notation of Section \ref{sub:The-bosonic-moduli-space}, the measure over the moduli space is
\[
\int_{\mathcal{M}_{G}^{0}\left(g,d\right)}=\sum_{\text{partitions }N}\:\prod_{j=1}^{p}\left(\sum_{m_{j}\in0,\ldots, dN_{j}-1}\int_{\mathcal{M}_{N_{j},m_{j}}^{g}}\int_{0}^{1}\frac{dx_{j}}{2\pi}\right) \ .
\]
The outermost sum runs over partitions of $N$ such that 
\[
\sum_{j=1}^{p}N_{j}=N\ ,\qquad N_{j}\ge1\ ,
\]
and 
\[
|\CW| = \prod_{l} n_{l}! \ ,
\]
 is the residual Weyl factor.

\end{itemize}

\subsection{\label{classical_contributions}Classical contributions}

Classical contributions corresponding to flat connections can not
come from the standard kinetic terms for either the matter or gauge
multiplet as these are $\delta$-exact. The same is true for the
renormalized $D$-terms in the effective action. In fact, the entire
field strength multiplet, and anything constructed out of it, will
vanish. The superpotential does not contribute because chiral multiplet
fields are all required to vanish. Any classical contribution would
have to come from $D$-terms constructed out of the vector superfield
$\CV$ or topological terms such as the discrete theta angles discussed
in \cite{Aharony:2013hda}.

A simple gauge invariant term constructed out of $\CV$, which appears
only for abelian factors of the gauge group $G$, is the Fayet-Iliopoulos
term. In superspace, this term is simply 
\[
\xi\CV\ .
\]
On curved space, the appropriate $D$-term is then
\footnote{The factor of $i$ appears because of our rotation of the integration
contour of $D$.
} 
\[
\xi\int\left(D-iV^{\mu}a_{\mu}\right)\ ,
\]
where 
\[
V_{\mu}=-\frac{1}{2}\nabla^{\nu}J_{\nu\mu}+\kappa K_{\mu}\ .
\]
$D$ vanishes on the moduli space and we can integrate by parts to
get rid of the term involving the complex structure. What remains
is 
\[
-i\xi\int\kappa\, K^{\mu}a_{\mu}=-i\xi a_{w}\int\kappa\ ,
\]
which depends on $a_\mu$ only through the holomorphic modulus $a_{w}$.

Alternatively, we could use the expression for $V_{\mu}$ as the dual
of the field strength $H_{\mu\nu\rho}$ for the supergravity two-form
$B_{\mu\nu}$. If $H$ is exact then there is a well-defined two-form
$B$ such that 
\[
V=\star H=\star dB\ ,
\]
and the FI term vanishes on the moduli space 
\[
\xi\int V^{\mu}a_{\mu}\propto\xi\int dB\wedge a=-\xi\int B\wedge F^{a}=0\ .
\]
To get a contribution, we must take some nontrivial $H\in H^{3}\left(M,\mathbb{Z}\right)$.
In fact following \cite{Closset:2013vra}, the contribution is due
to $W\in H^{1,2}\left(M\right)$ given by\footnote{The convention in \cite{Closset:2013vra} is that $K$ is anti-holomorphic,
hence the difference in the relevant cohomology class.
} 
\[
H=\frac{i}{2}dJ+W\ ,\qquad\partial W=0\ ,
\]
since as we have seen only $a_{w}$ contributes.

\subsection{Operators}

We can attempt to deform the indices by supersymmetric operators which
are annihilated by the supersymmetry transformation $\delta$. Since
the moduli space over which we integrate involves flat connections, it makes
sense to insert supersymmetric Wilson loops.

One would like to insert 
\[
\mathcal{W}_{\FR}=\text{Tr}_{\FR}\left[\mathcal{P}\exp\left(i\int\sqrt{g}\, K^{\mu}a_{\mu}\right)\right]\ ,
\]
where $\FR$ denotes a representation of the gauge group and $\mathcal{P}$
is the possibly ambiguous path ordering symbol. This is the would-be
analogue of the (supersymmetric) light-like Wilson loop in Minkowski
space. It is obviously BPS. The integration is over all of $M$, which is why we cannot make use of the usual path ordering. The operator is thus more like a smeared Wilson loop.

In the abelian case, this operator is also locally gauge invariant
because $K$ is co-closed (and $\mathcal{P}$ is unnecessary). However,
then we must worry about invariance under large gauge transformations.
When $M=S^{3}\times S^{1}$ the situation is better since the relevent large gauge
transformations are generated by functions into the group which ``wrap''
the $S^{1}$. One can ensure invariance by normalizing $K$ to have
unit holonomy around this cycle. In fact, the FI term is an example
of this construction. The ``charge'' of the Wilson loop is determined,
in that case, by the class of $H$ and the (quantized) FI parameter
$\xi$. The evaluation of the expectation value of a charge $q$ abelian
Wilson loop corresponds to an insertion, into the sum and integration
of the index, of 
\[
\exp\left(iqa_{w}^{U\left(1\right)}\right)\:,
\]
where $a_{w}^{U\left(1\right)}$ is the gauge $U\left(1\right)$-connection.

When the gauge group is non-abelian but simply-connected we must try
to fix the ambiguity in the definition of $\mathcal{W}_{\mathfrak{R}}$
in some other way. The main point of such an exercise would be to
recover, at the appropriate point in the complex structure moduli
space, the supersymmetric 3$d$ Wilson loop of the form
\[
\mathcal{W}_{\mathfrak{R}}^{\text{3}d}=\text{Tr}_{\FR}\left[\mathcal{P}\exp\left(\oint d\tau\left(ia_{\mu}\dot{x}^{\mu}+\sigma\left|\dot{x}\right|\right)\right)\right] \ ,
\]
where $\sigma$ is the scalar in the 3$d$ vector multiplet which
comes from reducing the 4$d$ vector field along the ``time'' circle and
$\dot{x}^{\mu}$ points along the fiber of $M_{3}\rightarrow\Sigma$.
When such an observable is well-defined, the result, at an arbitrary
complex structure, would be an insertion of the expression 
\[
\mathcal{W}_{\FR}=\sum_{\rho\in\FR}u_{\rho}\ ,
\]
where $\rho$ is a weight in $\FR$ and 
\[
u_{\rho}=\exp\left(2\pi i\rho\left(a_{w}\right)\right)\ .
\]

\section{Discussion}

We have arrived at an expression for the partition function of a 4$d$
$\mathcal{N}=1$ gauge theory, with a conserved $R$-symmetry current,
on $M\simeq S^{1}\times M_{3}$ and $M_{3}$ a nontrivial circle
bundle over a compact oriented Riemann surface $\Sigma$. The parameters
entering the partition function are split between 
\begin{enumerate}
\item Parameters and deformations of the theory

\begin{enumerate}
\item The gauge group $G$. 
\item The representation of the matter multiplets $\FR$. 
\item A set of admissible Fayet-Iliopoulos terms $\xi$, one for each independent
$U(1)$ factor in $G$. 
\item An element of the moduli space of flat connections on $M$ of the
flavor symmetry group $F$. 
\end{enumerate}
\item Parameters of $M$

\begin{enumerate}
\item The genus, $g$, of the underlying Riemann surface. 
\item The first Chern class, $d$, of the circle bundle whose total space
is $M_{3}$. 
\item A point in the complex structure moduli space on $M$ admitting a
holomorphic Killing vector $K$. This may include a discrete choice
in the case $g=0$. 
\item A choice of $W\in H^{1,2}\left(M\right)$. 
\end{enumerate}
\end{enumerate}
The final result is given in \eqref{eq:final_partition_function}.

Our derivation included some assumptions regarding the geometry of
$M$. Primarily, we assumed that $M$ admits a Hermitian metric with
a holomorphic Killing vector $K$ with holomorphic coefficients. $K$
is a complex linear combination of commuting generators embedded in
the compact isometry group for the metric on $M$. The coefficients
in this linear combination, with a fixed embedding, are the only metric
parameters to which the partition function is sensitive. We did not
provide a way of restricting the range of these parameters to the
space of admissible metrics or, indeed, prove that some finite range
exists. It would be interesting to find out what restrictions one
can put given the final result.

We have argued that gaugino zero modes are absent once $M_{3}$ is
restricted such that $d>0$. This argument relied on the assumption
that the action in the localizing term for the fiber $\hat{\varphi}_{0}$
is non-degenerate. We have not considered possible zero modes for
the fields in the chiral multiplets. Such zero modes could exist for
specific choices of $g,d$ and specific representations $\FR$. It is
sometimes possible to lift such zero modes using the flavor symmetry
deformations.

The final result for the partition function includes an integral over
the moduli space of flat $G$-connections on $M$. There exists an
alternative approach for integrating over the gauge zero modes using
abelianization (c.f. \cite{Blau:1995rs}). This was used in a very
similar context in \cite{Ohta:2012ev} and in \cite{Blau:2006gh}
and results in a greatly simplified integral. Another possibility
is the use of Higgs branch localization. This was implemented for
the superconformal index in \cite{Peelaers:2014ima,Yoshida:2014qwa}.

Exact results of the type presented here can be used in a number of
ways. A very common, but technically challenging, application is to
duality. Since the partition function is independent of the RG flow,
we can compare the result for putative IR dual theories. Partition
functions on manifolds of varying topologies can be used to study
refinements of duality involving global aspects of the theory as demonstrated
in \cite{Razamat:2013opa}. One can also use them to explore the mapping
of operators. As shown in \cite{DiPietro:2014bca}, the ``high temperature'' limit of the generalized indices, where the size of the $S^1$ factor shrinks, can be determined in terms of  the $a$ and $c$ type conformal anomalies of the theory.

Our results involve some of the simplest examples of manifolds of
the supersymmetry preserving type found in \cite{Dumitrescu:2012ha}.
The methods we have used can also be applied to manifolds where $K$
acts with finite isotropy groups. The fluctuation determinants can
still, in principle, be computed using the index theorem. Alternatively,
computing the partition function for a background preserving only
one supercharge would require integrating over the instanton moduli
space and would likely include having to deal with gaugino zero modes.

\acknowledgments
We would like to thank C.\,Closset, A.\,Kapustin, V.\,Pestun, S.\,Razamat, I.\,Shamir, Y.\,Tachikawa, B.\,Willett and E.\,Witten for valuable discussions.
TN is grateful to T.\,Kugo for email correspondence. 
The work of TN was supported by a JSPS postdoctoral fellowship for research abroad.
The work of IY was supported by NSF grant PHY-0756966.

\appendix

\section{Conventions}\label{ss:Convention}


\subsection{Spinors and Fierz identities}

Our convention is close to \cite{VanNieuwenhuizen:1981ae} except
the two-component part. The metric is given by $\delta_{mn}\,(m,n=1,\cdots,4)$
whose sign is $(++++)$ (Euclidean signature). The totally antisymmetric
Levi-Civita tensor $\varepsilon_{mnpq}$ has $\varepsilon_{1234}=1$.
The gamma matrices satisfy $\{\gamma_{m},\gamma_{n}\}=2\delta_{mn}$
and $\gamma_{5}=\gamma_{1}\gamma_{2}\gamma_{3}\gamma_{4}$ with $\gamma_{5}^{2}=1$.
All of them are hermitian and $4\times4$ matrices. Under the rotation
group $SO(4)=SU(2)_{L}\times SU(2)_{R}$, left- and right-handed spinors
$\zeta_{\alpha}$, $\tilde{\zeta}^{\dot{\alpha}}$ are $SU(2)_{L}$
and $SU(2)_{R}$ doublets, respectively. The four-component Dirac
spinor $\zeta$ can be decomposed into one left-handed and one right-handed
spinors by chirality projection operators 
\begin{align}
\zeta_{\alpha}=\left(\frac{1+\gamma_{5}}{2}\zeta\right)_{\alpha}\ ,\qquad\tilde{\zeta}^{\dot{\alpha}}=\left(\frac{1-\gamma_{5}}{2}\zeta\right)^{\dot{\alpha}}\ .
\end{align}
The indices are raised and lowered by multiplying anti-symmetric tensors
$\varepsilon^{\alpha\beta}$ and $\varepsilon_{\alpha\beta}$ ($\varepsilon^{12}=\varepsilon_{21}=1$)
from the left 
\begin{align}
\zeta^{\alpha}=\varepsilon^{\alpha\beta}\zeta_{\beta}\ ,\qquad\zeta_{\alpha}=\varepsilon_{\alpha\beta}\zeta^{\beta}\ .
\end{align}
An analogous convention holds for dotted spinors. With these spinors,
the inner products are defined by 
\begin{align}
\begin{aligned}\psi\chi & \equiv\psi^{\alpha}\chi_{\alpha}\ ,\\
\tilde{\psi}\tilde{\chi} & \equiv\tilde{\psi}_{\dot{\alpha}}\tilde{\chi}^{\dot{\alpha}}\ .
\end{aligned}
\end{align}

The hermitian conjugate of the spinors are defined by 
\begin{align}
\begin{aligned}(\zeta^{\dagger})^{\alpha}=\overline{(\zeta_{\alpha})}\ ,\qquad(\tilde{\zeta}^{\dagger})_{\dot{\alpha}}=\overline{(\tilde{\zeta}^{\dot{\alpha}})}\ .\end{aligned}
\end{align}
We also define the hermitian conjugation for two anti-commuting spinors
by 
\begin{align}
\overline{\zeta_{1}\zeta_{2}}\equiv\zeta_{2}^{\dagger}\zeta_{1}^{\dagger}\ .
\end{align}

We choose the representation of the gamma matrices as 
\begin{align}
\gamma_{m}=\left(\begin{array}{cc}
 & (\sigma^{m})_{\alpha\dot{\beta}}\\
(\bar{\sigma}^{m})^{\dot{\alpha}\beta}
\end{array}\right)\ ,\qquad\gamma_{5}=\left(\begin{array}{cc}
1\\
 & -1
\end{array}\right)\ ,
\end{align}
where the sigma matrices are 
\begin{align}
(\sigma^{m})_{\alpha\dot{\beta}}=(-i\vec{\sigma},1)\ ,\qquad(\bar{\sigma}^{m})^{\dot{\alpha}\beta}=\varepsilon^{\dot{\alpha}\dot{\gamma}}\varepsilon^{\beta\delta}(\sigma^{m})_{\delta\dot{\gamma}}=(\sigma^{m})^{\beta\dot{\alpha}}=(i\vec{\sigma},1)\ .\label{SigmaMatrices}
\end{align}
In this representation, the charge conjugation matrix $C$ for Dirac
spinors is given by 
\begin{align}
C=\left(\begin{array}{cc}
-\varepsilon^{\alpha\beta}\\
 & -\varepsilon_{\dot{\alpha}\dot{\beta}}
\end{array}\right)\ ,\qquad C^{-1}=\left(\begin{array}{cc}
-\varepsilon_{\alpha\beta}\\
 & -\varepsilon^{\dot{\alpha}\dot{\beta}}
\end{array}\right)\ .
\end{align}
The Dirac conjugate spinor defined by $\bar{\zeta}=\zeta^{T}C$ is
then decomposed to $\bar{\zeta}=(\zeta^{\alpha},\tilde{\zeta}_{\dot{\beta}})$.

The sigma matrices satisfy the identities 
\begin{align}
\begin{aligned}(\sigma^{m})_{\alpha\dot{\beta}}(\sigma_{m})_{\gamma\dot{\delta}} & =2\varepsilon_{\alpha\gamma}\varepsilon_{\dot{\beta}\dot{\delta}}\ ,\qquad(\sigma_{m})_{\alpha\dot{\beta}}(\sigma_{n})^{\alpha\dot{\beta}}=2g_{mn}\ ,\\
\sigma_{m}\bar{\sigma}_{n}+\sigma_{n}\bar{\sigma}_{m} & =2g_{mn}\ ,\qquad\bar{\sigma}_{m}\sigma_{n}+\bar{\sigma}_{n}\sigma_{m}=2g_{mn}\ ,\\
\sigma_{m}\bar{\sigma}_{n}\sigma_{l} & =\varepsilon_{mnlk}\sigma_{k}+\delta_{mn}\sigma_{l}+\delta_{nl}\sigma_{m}-\delta_{ml}\sigma_{n}\ ,\\
\bar{\sigma}_{m}\sigma_{n}\bar{\sigma}_{l} & =-\varepsilon_{mnlk}\bar{\sigma}_{k}+\delta_{mn}\bar{\sigma}_{l}+\delta_{nl}\bar{\sigma}_{m}-\delta_{ml}\bar{\sigma}_{n}\ ,
\end{aligned}
\end{align}
We can define the anti-symmetric matrices 
\begin{align}
\sigma_{mn}=\frac{1}{4}(\sigma_{m}\bar{\sigma}_{n}-\sigma_{n}\bar{\sigma}_{m})\ ,\qquad\bar{\sigma}_{mn}=\frac{1}{4}(\bar{\sigma}_{m}\sigma_{n}-\bar{\sigma}_{n}\sigma_{m})\ ,\label{SigmaAnti}
\end{align}
which satisfy 
\begin{align}
\sigma_{mn}=-\frac{1}{2}\varepsilon_{mnpq}\sigma^{pq}\ ,\qquad\bar{\sigma}_{mn}=\frac{1}{2}\varepsilon_{mnpq}\bar{\sigma}^{pq}\ ,
\end{align}
and 
\begin{align}
(\sigma_{mn})_{\alpha\beta}=(\sigma_{mn})_{\beta\alpha}\ ,\qquad(\bar{\sigma}_{mn})_{\dot{\alpha}\dot{\beta}}=(\bar{\sigma}_{mn})_{\dot{\beta}\dot{\alpha}}\ .
\end{align}
There are additional identities including the anti-symmetric matrices
\begin{align}
\begin{aligned} & \sigma^{m}\bar{\sigma}^{n}=2\sigma^{mn}+\delta^{mn}\ ,\qquad\bar{\sigma}^{m}\sigma^{n}=2\bar{\sigma}^{mn}+\delta^{mn}\ ,\qquad\sigma^{m}\bar{\sigma}_{mn}=\frac{3}{2}\sigma_{n}\ ,\qquad\bar{\sigma}^{m}\sigma_{mn}=\frac{3}{2}\bar{\sigma}_{n}\ ,\\
 & \text{tr}(\sigma_{mn}\sigma_{lk})=\frac{1}{2}(\varepsilon_{mnlk}+\delta_{mk}\delta_{nl}-\delta_{ml}\delta_{nk})\ ,\\
 & \text{tr}(\bar{\sigma}_{mn}\bar{\sigma}_{lk})=\frac{1}{2}(-\varepsilon_{mnlk}+\delta_{mk}\delta_{nl}-\delta_{ml}\delta_{nk})\ ,\\
 & \sigma_{mn}\sigma_{lk}=\frac{1}{4}(\varepsilon_{mnlk}+\delta_{mk}\delta_{nl}-\delta_{ml}\delta_{nk})-\frac{1}{2}(\delta_{ml}\sigma_{nk}+\delta_{nk}\sigma_{ml})+\frac{1}{2}(\delta_{mk}\sigma_{nl}+\delta_{nl}\sigma_{mk})\ ,\\
 & \bar{\sigma}_{mn}\bar{\sigma}_{lk}=\frac{1}{4}(-\varepsilon_{mnlk}+\delta_{mk}\delta_{nl}-\delta_{ml}\delta_{nk})-\frac{1}{2}(\delta_{ml}\bar{\sigma}_{nk}+\delta_{nk}\bar{\sigma}_{ml})+\frac{1}{2}(\delta_{mk}\bar{\sigma}_{nl}+\delta_{nl}\bar{\sigma}_{mk})\ ,
\end{aligned}
\end{align}

From $C\gamma_{m}C^{-1}=-\gamma_{\mu}^{T}$ and $\bar{\lambda}_{A}=\lambda^{B}C_{BA}$,
it follows 
\begin{align}
\begin{aligned} & \bar{\lambda}\gamma_{m_{1}}\cdots\gamma_{m_{n}}\chi=(-1)^{n}\bar{\chi}\gamma_{m_{n}}\cdots\gamma_{m_{1}}\lambda\ ,\\
 & \bar{\lambda}\gamma_{5}\chi=\bar{\chi}\gamma_{5}\lambda\ ,\qquad\bar{\lambda}\gamma_{5}\gamma_{m}\chi=\bar{\chi}\gamma_{5}\gamma_{m}\lambda\ .
\end{aligned}
\end{align}

It follows from the definition that the product of spinors satisfy
\begin{align}
\begin{aligned}\psi_{\alpha}\chi_{\beta} & =\psi_{\beta}\chi_{\alpha}+\varepsilon_{\alpha\beta}\psi\chi\ ,\\
\tilde{\psi}_{\dot{\alpha}}\tilde{\chi}_{\dot{\beta}} & =\tilde{\psi}_{\dot{\beta}}\tilde{\chi}_{\dot{\alpha}}-\varepsilon_{\dot{\alpha}\dot{\beta}}\tilde{\psi}\tilde{\chi}\ ,\\
\psi_{\alpha}\chi_{\beta} & =\frac{1}{2}\varepsilon_{\alpha\beta}\psi\chi-\frac{1}{2}(\sigma^{mn})_{\alpha\beta}\psi\sigma_{mn}\chi\ ,\\
\tilde{\psi}_{\dot{\alpha}}\tilde{\chi}_{\dot{\beta}} & =-\frac{1}{2}\varepsilon_{\dot{\alpha}\dot{\beta}}\tilde{\psi}\tilde{\chi}-\frac{1}{2}(\bar{\sigma}^{mn})_{\dot{\alpha}\dot{\beta}}\tilde{\psi}\bar{\sigma}_{mn}\bar{\chi}\ ,\\
\psi_{\alpha}\tilde{\chi}_{\dot{\alpha}} & =\frac{1}{2}\sigma_{\alpha\dot{\alpha}}^{m}\psi\sigma_{m}\tilde{\chi}\ ,
\end{aligned}
\end{align}
The Fierz identities we will often use are 
\begin{align}
\begin{aligned}(\psi_{1}\psi_{2})(\tilde{\psi}_{3}\tilde{\psi}_{4}) & =\frac{1}{2}(\psi_{1}\sigma^{m}\tilde{\psi}_{4})(\psi_{2}\sigma_{m}\tilde{\psi}_{3})\ ,\\
(\psi_{1}\psi_{2})(\psi_{3}\psi_{4}) & =-(\psi_{1}\psi_{3})(\psi_{2}\psi_{4})-(\psi_{1}\psi_{4})(\psi_{2}\psi_{3})\ ,\\
(\tilde{\psi}_{1}\tilde{\psi}_{2})(\tilde{\psi}_{3}\tilde{\psi}_{4}) & =-(\tilde{\psi}_{1}\tilde{\psi}_{3})(\tilde{\psi}_{2}\tilde{\psi}_{4})+(\tilde{\psi}_{1}\tilde{\psi}_{4})(\tilde{\psi}_{2}\tilde{\psi}_{3})\ .
\end{aligned}
\end{align}

\subsection{Spin connection and Lie derivatives}

The spin connection for a given vielbein is defined by 
\begin{align}
\omega_{\mu}^{~mn}(e)=e_{\nu}^{m}\nabla_{\mu}e^{\nu n}\ ,
\end{align}
and the Riemann tensor is given using the spin connection by 
\begin{align}
R_{\mu\nu}^{~~mn}(e)=\partial_{\mu}\omega_{\nu}^{~mn}-\partial_{\nu}\omega_{\mu}^{~mn}+\omega_{\mu}^{~ml}\omega_{\nu l}^{~n}-\omega_{\nu}^{~ml}\omega_{\mu l}^{~n}\ .
\end{align}
The Ricci scalar is then 
\begin{align}
R(e)=e_{m}^{\nu}e_{n}^{\mu}R_{\mu\nu}^{~~mn}(e)\ .
\end{align}
This convention yields a negative Ricci curvature for a round sphere.

The covariant derivatives for spinors are defined by 
\begin{align}
\nabla_{\mu}\zeta=\partial_{\mu}\zeta+\frac{1}{2}\omega_{\mu}^{~mn}\sigma_{mn}\zeta\ ,\qquad\nabla_{\mu}\tilde{\zeta}=\partial_{\mu}\tilde{\zeta}+\frac{1}{2}\omega_{\mu}^{~mn}\bar{\sigma}_{mn}\tilde{\zeta}\ .
\end{align}
The commutator of two covariant derivatives yields the integrability
conditions 
\begin{align}
[\nabla_{\mu},\nabla_{\nu}]\zeta=\frac{1}{2}R_{\mu\nu}^{~~mn}\sigma_{mn}\zeta\ ,\qquad[\nabla_{\mu},\nabla_{\nu}]\tilde{\zeta}=\frac{1}{2}R_{\mu\nu}^{~~mn}\bar{\sigma}_{mn}\tilde{\zeta}\ .
\end{align}
The Lie derivative of a spinor along a vector $X=X^{\mu}\partial_{\mu}$
is given by 
\begin{align}
\begin{aligned}\CL_{X}\zeta=X^{\mu}\nabla_{\mu}\zeta+\frac{1}{2}\nabla_{\mu}X_{\nu}\sigma^{\mu\nu}\zeta\ ,\\
\CL_{X}\tilde{\zeta}=X^{\mu}\nabla_{\mu}\tilde{\zeta}+\frac{1}{2}\nabla_{\mu}X_{\nu}\bar{\sigma}^{\mu\nu}\tilde{\zeta}\ .
\end{aligned}
\label{LieSpinor}
\end{align}

The Weyl tensor is define by 
\begin{align}
C_{\mu\nu\rho\sigma}=R_{\mu\nu\rho\sigma}+\frac{1}{2}\left(g_{\mu\rho}R_{\nu\sigma}+g_{\nu\sigma}R_{\mu\rho}-g_{\mu\sigma}R_{\nu\rho}-g_{\nu\rho}R_{\mu\sigma}\right)+\frac{R}{6}(g_{\mu\sigma}g_{\nu\rho}-g_{\mu\rho}g_{\nu\sigma})\ .
\end{align}

\subsection{Hermitian coordinates}

The holomorphic coordinates of $\BR^{4}$ are given by 
\begin{align}
z^1 = -x^2 + i x^1 \ , \qquad z^2 = x^4 + ix^3 \ ,
\end{align}
and the Levi-Civita tensor becomes $\varepsilon_{1\bar{1}2\bar{2}}=\frac{1}{4}$.
The sigma matrices in these coordinates are obtained from \eqref{SigmaMatrices}
by coordinate transformation 
\begin{align}\label{SigmaMatHermite}
\begin{aligned}\sigma_{1} & =-\bar{\sigma}_{1}=\left(\begin{array}{cc}
0~ &~0\\
-1 ~&~ 0
\end{array}\right)\ ,&\qquad\sigma_{\bar{1}}&=-\bar{\sigma}_{\bar{1}}=\left(\begin{array}{cc}
0~ & ~1\\
0~ &~ 0
\end{array}\right)\ ,\\
\sigma_{2} & =\bar{\sigma}_{\bar{2}}=\left(\begin{array}{cc}
0~ &~ 0\\
 0~&~ 1
\end{array}\right)\ ,&\qquad\sigma_{\bar{2}}&=\bar{\sigma}_{2}=\left(\begin{array}{cc}
1~ &~ 0\\
0~&~ 0
\end{array}\right)\ ,
\end{aligned}
\end{align}
and the anti-symmetric matrices \eqref{SigmaAnti} are given by 
\begin{align}
\begin{aligned}\sigma_{1\bar{1}} & =\bar{\sigma}_{1\bar{1}}=\sigma_{2\bar{2}}=-\bar{\sigma}_{2\bar{2}}= \frac{1}{4}\left(\begin{array}{cc}
-1~&~0\\
 0~&~ 1
\end{array}\right)\ ,\\
\sigma_{12} & =\bar{\sigma}_{1\bar{2}}=\frac{1}{2}\left(\begin{array}{cc}
0~ &~ 0\\
-1~ &~ 0
\end{array}\right)\ ,\qquad\sigma_{\bar{1} \bar{2}}=-\bar{\sigma}_{\bar{1}2}=\frac{1}{2}\left(\begin{array}{cc}
 0~&~ 1\\
0~&~ 0
\end{array}\right)\ .
\end{aligned}
\end{align}

\section{Special functions}
The Dedekind's eta function is defined by
\begin{align}\label{etaDef}
\eta (\tau) = q^{1/24} \prod_{n=1}^\infty (1-q^n) \ ,
\end{align}
where $q=e^{2\pi i \tau}$.

The Jacobi's theta functions are defined as follows:
\begin{align}\label{JacobiDef}
\begin{aligned}
\vartheta_1 (z|\tau) &= 2q^{1/8} \sin(\pi z)\prod_{n=1}^\infty (1-q^n) (1-y q^n) (1- y^{-1} q^{n}) \ , \\
\vartheta_2 (z|\tau) &= 2q^{1/8} \cos(\pi z)\prod_{n=1}^\infty (1-q^n) (1+y q^n) (1+ y^{-1} q^{n}) \ , \\
\vartheta_3 (z|\tau) &= \prod_{n=1}^\infty (1-q^n) (1+y q^{n-1/2}) (1+y^{-1} q^{n-1/2}) \ , \\
\vartheta_4 (z|\tau) &= \prod_{n=1}^\infty (1-q^n) (1-y q^{n-1/2}) (1-y^{-1} q^{n-1/2}) \ , 
\end{aligned}
\end{align}
with $q=e^{2\pi i \tau}$ and $y=e^{2\pi i z}$.
Also we define 
\[
\vartheta_{0}\left(z|\tau\right)
=\prod_{n=0}^{\infty}\left(1- y q^n\right)\left(1- y^{-1} q^n\right)  \ .
\]

There is a general formula for an infinite product
\[
\prod_{n\in\mathbb{Z}}\frac{n+a}{n+a+b}=e^{\pi ib}\frac{1-\exp\left(2\pi ia\right)}{1-\exp\left(2\pi i\left(a+b\right)\right)} \ .
\]
The elliptic gamma function is defined as (within some range of convergence)
\cite{felder2000elliptic}
\[
\Gamma\left(z,\tau,\sigma\right)=\prod_{l_{1},l_{2}\ge0}\frac{1-e^{2\pi i\left(\left(l_{1}+1\right)\tau+\left(l_{2}+1\right)\sigma-z\right)}}{1-e^{2\pi i\left(l_{1}\tau+l_{2}\sigma+z\right)}} \ ,
\]
or in an alternative region by 
\[
\Gamma\left(z,\tau,\sigma\right)=\exp\left(-\frac{i}{2}\sum_{j=0}^{\infty}\frac{\sin\left(\pi j\left(2z-\tau-\sigma\right)\right)}{j\sin\left(\pi j\tau\right)\sin\left(\pi j\sigma\right)}\right) \ ,
\]
hence
\[
\Gamma\left(z,\tau,\sigma\right)=e^{\pi i\left(\tau+\sigma-2z\right)}\prod_{n\in\mathbb{Z}|\, l_{1},l_{2}\ge0}\frac{n+\left(l_{1}+1\right)\tau+\left(l_{2}+1\right)\sigma-z}{n+l_{1}\tau+l_{2}\sigma+z} \ .
\]
Alternatively, using the variables
\[
u=e^{2\pi iz}\ ,\qquad p=e^{2\pi i\sigma}\ ,\qquad q=e^{2\pi i\tau} \ ,
\]
it is written as
\[
\Gamma\left(u;p,q\right)=\prod_{l_{1},l_{2}\ge0}\frac{1-u^{-1}p^{l_{1}+1}q^{l_{2}+1}}{1-u\,p^{l_{1}}q^{l_{2}}} \ .\label{DefEllipticGamma}
\]


\bibliographystyle{JHEP}
\bibliography{LocalizationOnM4}

\end{document}